\documentclass{article}

\usepackage[preprint]{neurips_2026}

\usepackage{microtype}
\usepackage{subcaption}
\usepackage{booktabs} 

\usepackage{algorithm}
\usepackage{algorithmic}

\usepackage{amsmath}
\usepackage{amssymb}
\usepackage{amsfonts}
\usepackage{mathtools}
\usepackage{amsthm}
\usepackage{bm}
\usepackage{bbm}

\theoremstyle{plain}

\theoremstyle{definition}

\theoremstyle{remark}

\usepackage{booktabs}
\usepackage{multirow}
\usepackage{tabularx}
\usepackage{array}
\usepackage[table]{xcolor}
\usepackage{soul}

\usepackage[T1]{fontenc}
\usepackage{nicefrac}
\usepackage{xcolor}
\usepackage{enumitem}
\usepackage{wrapfig}
\usepackage{multicol}
\usepackage{diagbox}
\usepackage{pifont}
\usepackage{pdfpages}
\usepackage{arydshln}
\usepackage{makecell}
\usepackage{adjustbox}
\usepackage{colortbl}

\definecolor{blue(pigment)}{rgb}{0.2, 0.2, 0.6}
\definecolor{aliceblue}{rgb}{0.85, 0.9, 1.0}
\definecolor{denim}{rgb}{0.08, 0.38, 0.74}
\definecolor{lapislazuli}{rgb}{0.15, 0.38, 0.61}
\definecolor{yaleblue}{rgb}{0.06, 0.3, 0.57}

\newcommand{\hlblue}[1]{\sethlcolor{aliceblue!60}\hl{#1}}

\usepackage[colorlinks=true, linkcolor=blue(pigment), citecolor=blue(pigment)]{hyperref}
\usepackage[capitalise,noabbrev]{cleveref}
\crefname{appendix}{Appendix}{Appendices}
\Crefname{appendix}{Appendix}{Appendices}
\crefformat{figure}{Figure~#2{\color{blue(pigment)}#1}#3}
\crefformat{table}{Table~#2{\color{blue(pigment)}#1}#3}
\crefformat{equation}{Equation~#2{\color{blue(pigment)}#1}#3}
\crefformat{section}{Section~#2{\color{blue(pigment)}#1}#3}
\crefformat{appendix}{Appendix~#2{\color{blue(pigment)}#1}#3}

\def\pz{{\phantom{0}}}

\title{A Systematic Evaluation of Co-folding Model Representations for Small-Molecule Learning}

\author{%
  Hyosoon Jang\vspace{0.25em}\quad Hyunjin Seo\quad Honghui Kim\quad Seonghyun Park\quad Taewon Kim \\ \textbf{Yunhui Jang\vspace{0.45em}\quad Sungsoo Ahn}\\
  KAIST\\
  \texttt{\{hyosoon.jang,sungsoo.ahn\}@kaist.ac.kr} \\
}

\begin{document}

\maketitle

\begin{abstract}
Small-molecule foundation models are typically pretrained on standalone molecular data, unlike vision and language models that often benefit from cross-modal or relational supervision. Protein-ligand co-folding provides a molecular analogue of such supervision by exposing models to atom-level ligand-protein interactions, raising the question of whether co-folding models can yield strong small-molecule representations. We study this question using Boltz2, a modern co-folding model, by transferring its atom-level ligand representations to standalone small-molecule tasks. Through systematic probing and distillation, we show that Boltz2 representations match or outperform existing models on the ADMET benchmark, accelerate molecular generative modeling, and improve sample efficiency in structure-guided ligand optimization. We further find that Boltz2 representations are complementary to those learned from conventional standalone molecular supervision, including 3D conformers, bioassay labels, and quantum-chemical properties. Finally, we extend representation alignment to reinforcement learning, showing that dense representation-level supervision can complement scalar rewards in molecular discovery. These results identify protein-ligand co-folding as a promising pretraining paradigm for small-molecule representation learning and position Boltz2 as a strong, off-the-shelf molecular foundation model.\footnote{Code: \url{https://github.com/hsjang0/boltz-as-FM}.}
\end{abstract}

\section{Introduction}

Large-scale representation learning has shown strong transferability across a wide range of downstream tasks, particularly in language~\citep{llama3, gpt4} and vision domains~\citep{vit, dinov2}. Foundation models with large-scale pretraining have become essential for achieving state-of-the-art performance in downstream applications, serving as powerful feature extractors~\citep{radford2021learning} for predictive tasks. Beyond predictive tasks, researchers have leveraged their representations to enhance generative modeling, including distillation into denoising models~\citep{yu2024representation} and use as latent embeddings for latent diffusion models~\citep{rae}. 

In the small-molecule domain, this paradigm has motivated foundation models that learn transferable atom-level representations for drug discovery tasks such as ADMET prediction~\citep{huang2021therapeutics}. Most existing models learn from molecules considered in isolation: each training example is a small molecule associated with molecule-level supervision such as bioassay labels and quantum-chemical properties~\citep{mendez2024mole,klaser2024minimol,kim2024quantum,zhou2023unimol,li2023knowledge,sypetkowski2024scalability}. Although these models differ in architecture and objective, they share a common design choice: the small molecule itself is the primary object of representation learning, without explicit modeling of an interacting molecular partner.



This contrasts with vision and language, where strong representations often arise from supervision beyond standalone data, such as cross-modal or relational signals~\citep{radford2021learning,jia2021scaling,zhai2022lit}. We argue that the molecular domain has an analogous but underexplored source of supervision: protein-ligand co-folding structures~\citep{burley2021rcsb}, in which a small molecule is observed within an interacting protein context. 

We hypothesize that protein-ligand structures provide richer supervision than standalone molecular data by exposing atom-level interactions in bound conformations, including hydrogen bonding and electrostatic interactions. Timely, modern protein models, such as AlphaFold3~\citep{abramson2024accurate} and its open-weight counterpart Boltz~\citep{wohlwend2024boltz1,passaro2025boltz2}, have transitioned from residue-level to atom-level representation modeling for protein-ligand structure prediction. Although designed for structure prediction, we can transfer their atomistic representations to small-molecule tasks. We therefore ask:
\begin{center}
\noindent\fcolorbox{black}{gray!15}{
\parbox{0.92\linewidth}{
\centering
\textit{Can protein-ligand co-folding models provide strong atom-level representations for standalone small-molecule tasks?}
}
}
\end{center}

Although repurposing protein–ligand co-folding models for small-molecule representation learning is conceptually straightforward, it has not been systematically studied. This gap is important because the broader utility of co-folding models remains debated: recent studies suggest that models such as AlphaFold3 may rely heavily on MSA-derived evolutionary patterns rather than a genuine atom-level understanding of molecular interactions \citep{stahl2023protein,brotzakis2025alphafold}. Establishing this baseline therefore allows us to test whether the interaction-aware supervision learned by modern co-folding models transfers beyond their original structure-prediction objective.

\begin{figure}
\centering
\centering \includegraphics[width=0.85\linewidth]{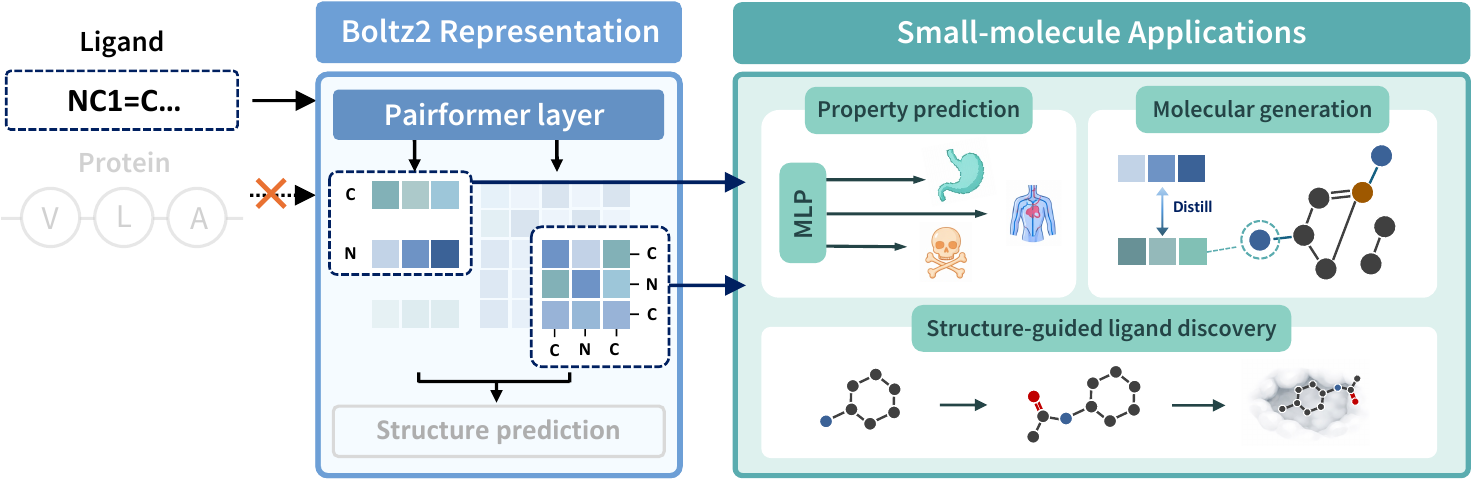}
\caption{\textbf{Boltz2 as an atom-level molecular foundation model.} We repurpose Boltz2, trained for protein–ligand co-folding, as a small-molecule representation model by leveraging its ligand representations and evaluating them on downstream small-molecule tasks.}\label{fig:boltz_as_fm}
\vspace{-.1in}
\end{figure}

\textbf{Contribution.} In this work, we investigate the potential of a co-folding model, Boltz2, as a source of representations for small-molecule tasks, bridging cutting-edge protein-centric models and atom-level representation learning for small molecules. As illustrated in \Cref{fig:boltz_as_fm}, we extract atom-level representations from Boltz2 for a given small molecule and evaluate their expressiveness across a broad range of small-molecule downstream tasks. Notably, our work is the first to show that informative molecular representations can improve molecular generation and optimization.

Through systematic experiments, we identify four key findings that position modern co-folding models, in particular Boltz2, as a strong baseline for small-molecule foundation models.
\begin{itemize}[topsep=-1.0pt,itemsep=1.0pt,leftmargin=3.5mm]
\item \textbf{Predictive transfer (\Cref{sec:admet})}: Boltz2 representations match or outperform specialized small-molecule foundation models on the extensive TDC ADMET benchmark.
\item \textbf{Generative transfer (\Cref{sec:mol_gen}):} Boltz2 representations improve molecular generative models with representation alignment-based distillation~\citep{yu2024representation}.
\item \textbf{Optimization transfer (\Cref{sec:structure_guided}):} Boltz2 representations enhance sample efficiency for designing ligands to maximize Boltz2's binding affinity score using a new distillation method. 
\item \textbf{Representation analysis (\Cref{sec:ablation}):} Boltz2 yields a representation space that complements existing molecular models. We further analyze the effects of protein context and layer depth.
\end{itemize}
Finally, we provide actionable insights for the molecular machine learning community. First, we extend representation alignment to online reinforcement learning for molecular discovery, showing how dense signals from the reward model can be used to improve sample efficiency beyond scalar rewards alone. Second, we show that combining weakly aligned representations, e.g., Boltz2 with MolE \citep{mendez2024mole}, outperforms more strongly aligned combinations, e.g., Boltz2 with MiniMol \citep{klaser2024minimol}, suggesting a practical strategy for representation fusion.

\section{Background}
\label{sec:background}

\subsection{Atom-level Representation Learning for Small-molecule Domain}

Atom-level representation learning has been widely explored to support diverse molecular property prediction tasks, such as ADMET profiles, under minimal task-specific supervision. These approaches adopt large-scale atom-level pretraining to learn transferable representations that capture atomic interactions within molecules. Building on this paradigm, existing methods investigate a range of pretraining objectives using a large collection of small molecules. MolE~\citep{mendez2024mole} uses masked atom prediction, MiniMol~\citep{klaser2024minimol} leverages large-scale bioassay supervision, QIP~\citep{kim2024quantum} incorporates quantum-chemical property regression to capture electronic structure information relevant to molecular properties, and UniMol~\citep{zhou2023unimol,ji2024exploring} learns representations through 3D molecular structure prediction. Other approaches incorporate chemical knowledge or scalability: KPGT~\citep{li2023knowledge} injects knowledge-guided objectives based on molecular structure and functional groups, UniQSAR \citep{gao2023uniqsarautomltoolmolecular} advances ADMET prediction using UniMol, and MolGPS~\citep{sypetkowski2024scalability} studies scalable pretraining across datasets and model sizes. 

\subsection{Atom-level Representations in Modern Co-folding Models}

Cutting-edge protein co-folding models operate at atomistic granularity and explicitly incorporate ligands, enabling the prediction of protein-ligand complex structures. Representative models such as Boltz~\citep{wohlwend2024boltz1, passaro2025boltz2} and AlphaFold3~\citep{abramson2024accurate} are trained to predict 3D structures of protein-ligand complexes at atomic resolution for both proteins and ligands.
These models learn joint representations of ligand atoms and protein residues under supervision from protein-ligand complex structures. Here, ligand atom representations are trained from their 3D conformations relative to surrounding protein atoms, implicitly encoding geometric and chemical traits of the bound state. 

By learning representations at atomistic resolution within protein-ligand complexes, co-folding models naturally bridge protein-centric and small-molecule-centric representation learning. Although their training objectives are defined by protein-ligand structure prediction, the ligand representations capture atom-level interaction patterns, such as hydrogen bonding and electrostatic interactions, that have the potential to support standalone small-molecule tasks.

\textbf{Boltz2 representation.} We primarily evaluate Boltz2~\citep{passaro2025boltz2} in this paper as an atom-level foundation model for small molecules. The model scale is 1B, consisting of a $64$-layer Pairformer trunk that produces pair and single representations over protein residues and ligand atoms for predicting 3D protein-ligand complex structures. Pair representations with $128$ dimensions encode residue-residue, residue-atom, and atom-atom interactions, while single representations with $384$ dimensions capture per-entity features updated via the pair representations. For standalone small-molecule tasks (\Cref{sec:admet,sec:mol_gen}), we omit protein inputs and consider ligands only. For structure-guided ligand discovery (\Cref{sec:structure_guided}), we retain protein inputs and use Boltz2 in its native protein-ligand setting. 

\newcolumntype{C}[1]{>{\centering\arraybackslash}p{#1}}\newcolumntype{L}[1]{>{\raggedright\arraybackslash}p{#1}}

\begin{table*}[t]
\centering
\caption{\textbf{Results on ADMET property prediction benchmark.}
$^{\dagger}$Other denotes the highest score reported on the TDC ADMET leaderboards as of Jan 2026, excluding the baselines in the table. {We evaluate the 1B-scale UniMol2 model.} Mini. denotes MiniMol. V., S., H., and M. denote Veith, Substrate, Hepatocyte, and Microsome, respectively. The results are averaged over five random seeds. \textbf{Bold} numbers indicate the best performance, while \underline{underlined} numbers indicate the best performance without ensemble. Overall, \hlblue{Boltz2} shows competitive performance compared to baselines. $^{\ast}$This performance can be improved by \textit{incorporating relevant protein} as input to Boltz2 (\Cref{tab:metabolism_protein}). See \Cref{tab:admet_boltz_only} in \Cref{supp:full_admet} for the standard deviations.}
\label{tab:admet_highlevel}
\scalebox{0.84}{
\begin{tabular}{
L{2.2cm}
C{0.7cm}
C{0.75cm}
C{0.75cm}
C{0.75cm}
C{0.75cm}
C{0.94cm}                         
>{\columncolor{aliceblue!60}}C{0.88cm} 
C{1.10cm}
C{1.10cm}
>{\columncolor{aliceblue!60}}C{1.20cm}
}
\toprule
\multicolumn{2}{@{}c@{}}{} &
\multicolumn{6}{@{}c@{}}{\footnotesize\textbf{w/ Pretraining}} &
\multicolumn{3}{@{}c@{}}{\footnotesize\textbf{w/ Representation Ensembling}} \\
\cmidrule(lr){3-8}
\cmidrule(lr){9-11}
{\footnotesize\textbf{Dataset}} &
{\footnotesize\textbf{$^{\dagger}$Other}} &
{\footnotesize\textbf{MolE}} &
{\footnotesize\textbf{KPGT}} &
{\footnotesize\textbf{Mini.}} &
{\footnotesize\textbf{QIP}} &
{\footnotesize\textbf{UniMol2}} &
{\footnotesize\textbf{Boltz2}} &
{\footnotesize\textbf{MolGPS}} &
{\footnotesize\textbf{UniQSAR}} &
{\footnotesize\textbf{Boltz2}$^{\mathrm{Mini.}}$} \\
\midrule
\hline

\rowcolor{gray!12}\multicolumn{11}{l}{\textit{Absorption}} \\
Caco2 $\downarrow$          & \underline{\textbf{0.26}}  & 0.31  & 0.28  & 0.35  & {0.27}  & 0.41  & 0.30  & 0.29  & 0.27  & 0.30 \\
HIA $\uparrow$              & \underline{\textbf{0.99}}  & 0.96  & 0.98  & \underline{\textbf{0.99}}  & \underline{\textbf{0.99}}  & 0.87  & \underline{\textbf{0.99}}  & 0.98  & \textbf{0.99}  & \textbf{0.99} \\
Pgp $\uparrow$              & \underline{0.94}  & 0.92  & \underline{0.94}  & \underline{0.94}  & 0.93  & 0.90  & 0.93  & \textbf{0.95}  & 0.93  & 0.93 \\
Bioavail. $\uparrow$        & \underline{0.75}  & 0.65  & \underline{0.75}  & 0.69  & 0.73  & 0.64  & \underline{0.75}  & 0.70  & 0.73  & \textbf{0.77} \\
Lipophilicity $\downarrow$  & 0.47  & 0.47  & {0.45}  & 0.46  & \underline{0.44}  & 0.61  & {0.45}  & \textbf{0.39}  & 0.42  & 0.41 \\
Solubility $\downarrow$     & 0.76  & 0.79  & 0.71  & 0.74  & 0.70  & 0.81  & \underline{0.66}  & 0.68  & 0.68  & \textbf{0.64} \\

\hline
\rowcolor{gray!12}\multicolumn{11}{l}{\textit{Distribution}} \\
BBB $\uparrow$              & 0.92  & 0.90  & 0.91  & {0.92}  & 0.90  & 0.86  & \underline{0.93}  & \textbf{0.94}  & 0.93  & \textbf{0.94} \\
PPBR $\downarrow$           & 7.44  & 8.07  & 7.68  & 7.70  & \underline{7.36}  & 8.95  & {7.65}  & \textbf{6.46}  & 7.53  & 7.59 \\
VDss $\uparrow$             & 0.71  & 0.65  & {0.63}  & 0.54  & 0.61  & 0.70  & \underline{0.74}  & 0.65  & 0.73  & \textbf{0.75} \\

\hline
\rowcolor{gray!12}\multicolumn{11}{l}{\textit{Metabolism}} \\
CYP2C9 V. $\uparrow$        & \underline{\textbf{0.86}}  & 0.80  & 0.80  & 0.82  & 0.78  & 0.77  & \,\,{0.82}$^{\ast}$  & 0.84  & 0.80  & \textbf{0.86} \\
CYP2D6 V. $\uparrow$        & \underline{\textbf{0.79}}  & 0.68  & 0.72  & 0.72  & 0.66  & 0.62  & \,\,0.69$^{\ast}$  & 0.75  & 0.74  & 0.72 \\
CYP3A4 V. $\uparrow$        & \underline{\textbf{0.92}}  & 0.87  & 0.89  & 0.88  & 0.87  & 0.81  & \,\,0.85$^{\ast}$  & 0.90  & 0.89  & 0.88 \\
CYP2C9 S. $\uparrow$        & 0.44  & 0.45  & 0.45  & 0.48  & \underline{\textbf{0.52}}  & 0.33  & 0.36  & 0.46  & 0.45  & 0.36 \\
CYP2D6 S. $\uparrow$        & \underline{\textbf{0.74}}  & 0.70  & \underline{\textbf{0.74}}  & 0.73  & 0.67  & 0.44  & 0.52  & 0.71  & 0.72  & 0.51 \\
CYP3A4 S. $\uparrow$        & 0.67  & 0.67  & \underline{\textbf{0.73}}  & 0.64  & 0.62  & 0.58  & 0.62  & {0.68}  & 0.65  & 0.60 \\

\hline
\rowcolor{gray!12}\multicolumn{11}{l}{\textit{Excretion}} \\
Half Life $\uparrow$        & 0.58  & 0.55  & 0.53  & 0.50  & 0.53  & 0.48  & \underline{0.62}  & 0.63  & 0.61  & \textbf{0.65} \\
Clearance H. $\uparrow$     & 0.54  & 0.38  & 0.42  & 0.45  & 0.50  & 0.45  & \underline{\textbf{0.62}}  & 0.57  & 0.49  & {0.60} \\
Clearance M. $\uparrow$     & 0.63  & 0.61  & 0.64  & 0.63  & \underline{\textbf{0.66}}  & 0.62  & 0.61  & 0.63  & 0.65  & {0.65} \\

\hline
\rowcolor{gray!12}\multicolumn{11}{l}{\textit{Toxicity}} \\
LD50 $\downarrow$           & 0.55  & 0.82  & 0.55  & 0.59  & 0.56  & 0.50  & \underline{\textbf{0.40}}  & 0.56  & 0.55  & \textbf{0.40} \\
hERG $\uparrow$             & {\textbf{0.88}}  & 0.81  & 0.85  & 0.85  & 0.82  & 0.75  & 0.86  & 0.86  & 0.86  & 0.86 \\
AMES $\uparrow$             & 0.87  & 0.88  & 0.87  & 0.85  & 0.86  & 0.88  & \underline{\textbf{0.91}}  & 0.86  & 0.88  & \textbf{0.91} \\
DILI $\uparrow$             & 0.93  & 0.58  & 0.93  & \underline{\textbf{0.96}}  & 0.89  & 0.83  & 0.89  & 0.94  & 0.94  & 0.87 \\

\hline
\multicolumn{2}{c}{\textbf{\# 1st} {\underline{\small(\# 1st w/o Ensembling)}}\pz}
& 0 (0)  & 2 (4)  & 2 (3)  & 3 (5)  & 0 (0)  & 4 (9)  & 4  & 1  & 9 \\
\multicolumn{2}{c}{{\# 2nd {\small(\# 2nd w/o Ensembling)}}}
& 1 (2)  & 4 (7)  & 3 (5)  & 2 (4)  & 1 (1)  & 5 (4)  & 8  & 8  & 4 \\
\bottomrule
\end{tabular}}
\vspace{-.1in}
\end{table*}

\begin{figure*}
\setcounter{figure}{1}  
\centering
\centering \includegraphics[width=0.99\linewidth]{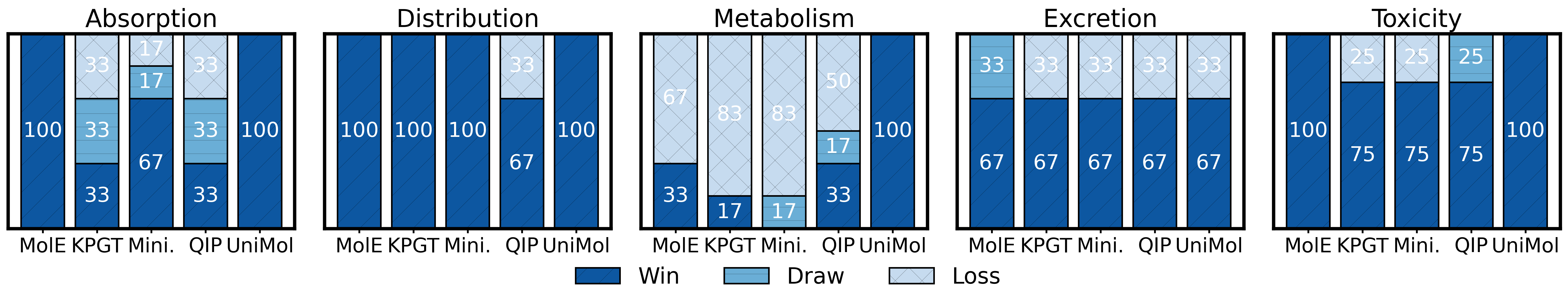}
\vspace{-.05in}
\caption{\textbf{Boltz2 vs. ADMET specialist foundation models.} {As illustrated,} Boltz2 shows competitive performance compared with existing foundation models specialized for ADMET property prediction in four of the five domains, despite not being designed for ADMET tasks.}\label{fig:admet_wdl}
\vspace{-.15in}
\end{figure*}

\section{ADMET Property Prediction}\label{sec:admet}

To assess Boltz2 as an atom-level representation model for small molecules, we first consider ADMET property prediction, a standard benchmark in the literature~\citep{mendez2024mole,klaser2024minimol,kim2024quantum,li2023knowledge,sypetkowski2024scalability}. This benchmark covers absorption, distribution, metabolism, excretion, and toxicity prediction tasks, which depend on both global molecular structure and local atomic interactions.

\subsection{Experimental Setup}

\textbf{Datasets.} We conduct experiments on the Therapeutics Data Commons (TDC) ADMET benchmark datasets \citep{huang2021therapeutics}, which consist of 22 ADMET property prediction tasks for small molecules represented as SMILES \citep{weininger1988smiles}. We follow the standard benchmark protocol, including data splits, evaluation metrics, and random seeds, where dataset statistics and benchmark settings are provided in \Cref{supp:admet_statistics}.

\textbf{Implementation details.} We apply probing to Boltz2 molecular representations for property prediction. We provide SMILES strings of molecules to the ligand modality of Boltz2, and protein sequence inputs are omitted. Then, we concatenate atom-wise pair representations from the $\{16, 32, 48, 64\}$-th layers of the $64$-layer Pairformer trunk. The pooled representation is fed into a probing network for target label prediction. We provide implementation details in \Cref{supp:admet_details}.\footnote{We evaluate the contribution of representations from each layer through ablation studies in \Cref{tab:layerwise}.}

\textbf{Baselines.} We consider six foundation models specialized for ADMET property prediction (See comparison with three more foundation models in \Cref{tab:boltz_vs_molgps_1b} of \Cref{app:scale}):
\begin{itemize}[topsep=-1.0pt,itemsep=1.0pt,leftmargin=3.5mm]
    \item \textbf{MolE}~\citep{mendez2024mole}: pretrained via masked language modeling on atom tokens with auxiliary losses to predict molecular properties and fingerprints.
    \item \textbf{KPGT}~\citep{li2023knowledge}: pretrained with knowledge-guided objectives that enforce consistency with chemical structures, functional groups, and expert-defined rules.
    \item \textbf{MiniMol}~\citep{klaser2024minimol}: pretrained to predict various labels spanning quantum chemistry, biological assays, and transcriptomic responses.
    \item \textbf{QIP}~\citep{kim2024quantum}: pretrained to approximate electronic structure properties such as orbital interactions.
    \item \textbf{UniMol2}~\citep{zhou2023unimol,ji2024exploring}: pretrained with a 1B Pairformer to predict standalone 3D molecular structures. Since no official results are available, we evaluate it under the same settings as ours.
    \item \textbf{MolGPS}~\citep{sypetkowski2024scalability}: ensembles three 1B-scale pretrained models, i.e., 3B scale, trained to predict bioassay labels or quantum properties.
    \item \textbf{UniQSAR}~\citep{gao2023uniqsarautomltoolmolecular}: advances UniMol \citep{zhou2023unimol}, which is pretrained to predict 3D structures of standalone molecules, by ensembling with various pretrained models to solve ADMET.
\end{itemize}
Among these, MolGPS and UniQSAR are ensemble-based methods over multiple pretrained models. For a direct comparison, we combine Boltz2 representations with those from MiniMol (Boltz2$^\text{Mini.}$).

\subsection{Results} 

As reported in \Cref{tab:admet_highlevel}, Boltz2 representations exhibit competitive performance. Among non-ensemble methods, Boltz2 achieves the best average performance on 9 of the 22 tasks. When evaluated under ensemble settings, Boltz2$^\text{Mini.}$ likewise attains the best results on 9 tasks. We further summarize win, draw, loss statistics in \Cref{fig:admet_wdl}. Boltz2 outperforms existing models on the majority of tasks. These results indicate that Boltz2 representations transfer effectively to standalone small-molecule property prediction, despite having no standalone small-molecule pretraining. 

To isolate the contribution of co-folding-based training, we clarify whether Boltz2's performance can be explained by 1B Pairformer architecture or data scale. UniMol2 provides a controlled comparison, as it also uses a 1B Pairformer architecture but is pretrained for \textit{standalone 3D molecular structure prediction}. Boltz2 significantly outperforms UniMol2, despite being pretrained on 750K complexes compared to UniMol2's 884M standalone molecules. Boltz2 also outperforms the three 1B-scale MolGPS variants (Appendix~\ref{app:scale}). These results suggest that co-folding supervision provides signal that is not recovered by architecture choice or scaling parameters and data alone.

In the analysis section (\Cref{sec:ablation}), we also analyze how Boltz2 provides representations that are complementary to those of existing molecular foundation models beyond the observed improvement from ensembling (Boltz2$^{\text{Mini.}}$). We also find that the Boltz2 performance can be further improved by incorporating relevant protein context, as shown in \Cref{tab:metabolism_protein} of \Cref{sec:ablation}.

Intriguingly, both Boltz2 and UniMol2 underperform on metabolism substrate prediction. On the baseline side, the gap stems from leakage in the bioassay pretraining datasets~\citep{sypetkowski2024scalability,koleiev2026critical}. On the Boltz2 side, this may reflect reported limitations in capturing local mechanisms~\citep{masters2025investigating}, which are central to substrate prediction. However, UniMol2 also shows the same weakness, suggesting a broader limitation of 3D structure-based learning rather than a Boltz2-specific issue (discussed in \Cref{sec:discussion}).

\section{Molecular Generation} \label{sec:mol_gen}

\begin{wrapfigure}{r}{0.5\linewidth}
\centering
\vspace{-.15in}
\centering \includegraphics[width=\linewidth]{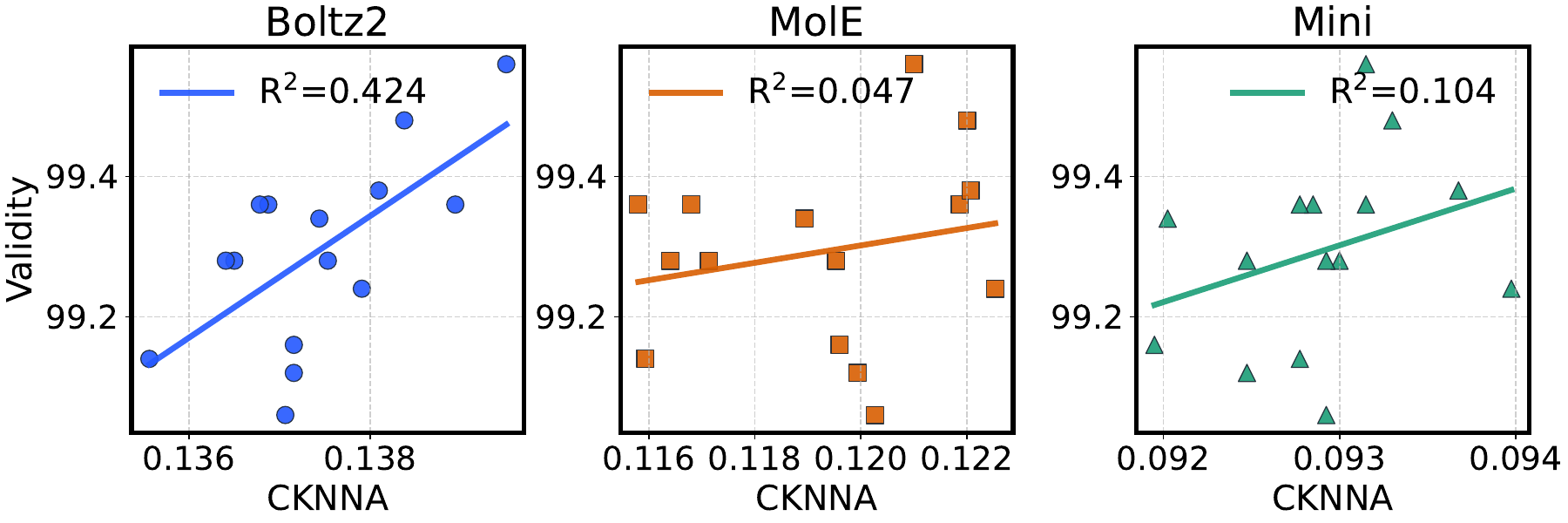}
\vspace{-.15in}
\caption{\textbf{Representation alignment vs. generation quality.} Stronger alignment with Boltz2 correlates with higher generation quality.}\label{fig:gen_cknna}
\vspace{-.1in}
\end{wrapfigure}

We next evaluate the quality of Boltz2 representations on small-molecule generation tasks. This experiment is motivated by recent work showing that representations from high-quality foundation models can improve the training of generative models via distillation \citep{yu2024representation}. Following this approach, we distill Boltz2 representations into state-of-the-art molecular generative models and assess their representation quality via acceleration of generative training.

\subsection{Observational Experiment} 

We first show that well-trained molecular generative models learn molecular representations that closely align with those produced by Boltz2. This alignment supports the use of Boltz2 representations as an additional training signal when training generative models from scratch \citep{yu2024representation}.

To be specific, following prior work~\citep{yu2024representation}, we measure the representation alignment between Boltz2 and a state-of-the-art molecular generation model, GruM~\citep{jo2023graph}. Specifically, we adopt CKNNA~\citep{huh2024platonic}, which quantifies the alignment between representations from molecular foundation models and generative models. Details of CKNNA are provided in \Cref{supp:cknna}.

We report alignment scores and generation quality of generative models in \Cref{fig:gen_cknna}. One can see that stronger alignment between Boltz2 and the generative models correlates with improved molecular generation quality. Building on this observation, we apply distillation to explicitly enforce representation alignment with Boltz2 during the training of molecular generative models.

\subsection{Experimental Setup}

\textbf{Datasets.} We conduct molecular generation experiments on the widely used ZINC250k dataset~\citep{jo2022score,vignac2022digress,jangsimple,kong2023autoregressive,jo2023graph,seo2025learning}. We train generative models on this dataset and evaluate their performance on $10{,}000$ generated molecules following prior work.

\textbf{Implementation details.} We adopt a representation alignment-based distillation \citep{yu2024representation} to existing diffusion-based molecular generative models, specifically GruM~\citep{jo2023graph}. We parameterize the denoising model of GruM with a four-layer Pairformer that outputs single and pair representations as hidden representations for denoising atom and bond types, respectively.

\begin{wrapfigure}{r}{0.5\linewidth}
\centering
\vspace{-.15in}
\centering \includegraphics[width=\linewidth]{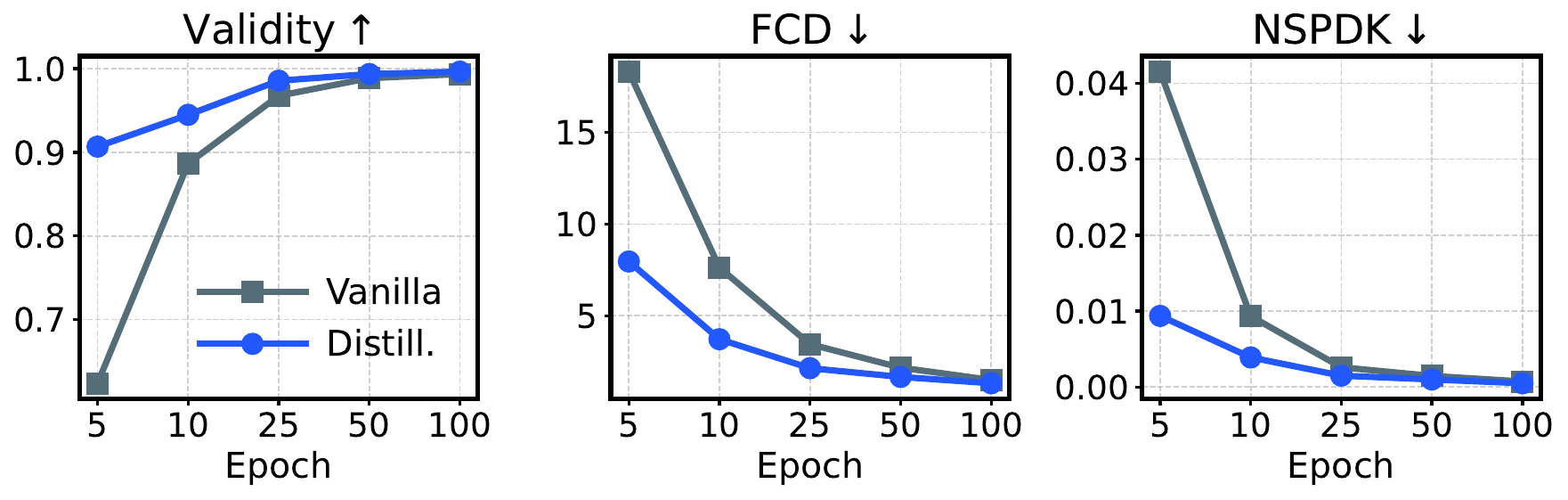}
\vspace{-.2in}
\caption{\textbf{2$\times$ faster training using Boltz2.} Representation alignment with Boltz2 makes generative model training 2$\times$ faster.}\label{fig:acceleration_gen}
\vspace{-.1in}
\end{wrapfigure}

We distill Boltz2 representations into GruM via representation alignment by training the denoising model to maximize the cosine similarity between its hidden representations of noisy molecules and the corresponding Boltz2 representations of clean molecules. We apply the alignment loss at the middle layer of the denoising model, aligning its single and pair representations with Boltz2. Implementation details are provided in \Cref{supp:molgen_details}.

\textbf{Baselines.} We consider five state-of-the-art diffusion-based graph generative models: original
GruM~\citep{jo2023graph}, GBD~\citep{liu2025gbd}, DeFoG~\citep{madeira2025defog},
TopBF~\citep{topbf2025}, and marginal SID~\citep{boget2025simple}. We also apply
representation alignment to GruM$^{\ast}$ using molecular representations from MolE,
KPGT, MiniMol, QIP, and UniMol2 to compare Boltz2-based alignment
with alternative representation sources.

\newcolumntype{L}[1]{>{\raggedright\arraybackslash}p{#1}}
\newcolumntype{C}[1]{>{\centering\arraybackslash}p{#1}}

\begin{table*}[t]
\caption{\textbf{Results on unconditional molecular generation.} GruM$^{\ast}$ denotes GruM parameterized with a Pairformer. The results are averaged over three random seeds. \textbf{Bold} numbers indicate the best performance. Representation alignment with \hlblue{Boltz2} accelerates the training of molecular generative models (\Cref{fig:acceleration_gen}, 2$\times$ faster training) to produce higher-quality molecules compared to the baselines. We report the standard deviation in \Cref{tab:full_gen} of \Cref{appx:full_gen}.}
\vspace{-.05in}
\label{tab:generation_quality}
\centering
\scalebox{0.86}{
\begin{tabular}{l c c c c c c c c }
\toprule
{\pz\pz\textbf{Method}} &
{\textbf{Valid $\uparrow$}} &
{\textbf{FCD $\downarrow$}} &
{\textbf{NSPDK $\downarrow$}} &
{\textbf{Novel $\uparrow$}} &
{\textbf{Unique $\uparrow$}} &
{\textbf{Scaffold $\uparrow$}} &
{\textbf{Fragment $\uparrow$}} &
{\textbf{SNN $\uparrow$}} \\
\hline
\rowcolor{gray!12}\multicolumn{9}{l}{\textit{State-of-the-art diffusion-based graph generative models}} \\
\textbf{GruM}                  & 98.65 & \pz2.26  & 0.0015 & 99.98 & 99.97 & 0.5299 & -- & -- \\
\textbf{GBD}                   & 97.87 & \pz2.25  & 0.0018 & -- & -- & 0.5042 & -- & -- \\
\textbf{DeFoG}                 & 99.22 & \pz1.42 & 0.0008 & -- & 99.99 & \textbf{0.5903} & -- & -- \\
\textbf{TopBF}                 & 99.37 & \pz1.39 & 0.0008 & -- & -- & 0.5372 & -- & -- \\
\textbf{Marg. SID}                   & 99.50 & \pz2.01 & 0.0021 & -- & -- & -- & -- & -- \\
{\textbf{GruM}$^{\ast}$}
& 99.36 & \pz1.49 & 0.0007& 99.99 & \textbf{100.00}\pz & 0.4923 & 0.9852 & 0.3697 \\
\hline
\rowcolor{gray!12}\multicolumn{9}{l}{\textit{Applying representation alignment-based distillation to GruM$^{\ast}$}} \\

\textbf{MolE}
& 99.51 & \pz1.41 & 0.0006 & \textbf{100.00}\pz & \textbf{100.00}\pz & 0.4932 & 0.9864 & {0.3737} \\

\textbf{KPGT}
& 99.37 & \pz1.46 & 0.0006 & \pz{99.99}\pz & \textbf{100.00}\pz & 0.4800 & 0.9860 & 0.3702 \\

\textbf{Mini.}
& 99.48 & \pz1.44 & 0.0007 & 99.99 & \textbf{100.00}\pz & 0.4812 & 0.9856 & 0.3710 \\

\textbf{QIP}
& 99.37 & \pz1.43 & 0.0007 & \textbf{100.00}\pz & \textbf{100.00}\pz & 0.5242 & 0.9864 & 0.3718 \\

\textbf{UniMol2} & 99.60 & \pz1.40 & 0.0006 & \textbf{100.00}\pz &  \textbf{100.00}\pz & 0.4864 & 0.9867 & 0.3721 \\

\rowcolor{aliceblue!60}
\textbf{Boltz2}
& \textbf{99.65} & \pz\textbf{1.31} & \textbf{0.0005} & \textbf{100.00}\pz & \textbf{100.00}\pz & 0.5064 & \textbf{0.9881} & \textbf{0.3766} \\
\bottomrule
\end{tabular}}
\vspace{-.15in}
\end{table*}

\textbf{Metrics.} In this experiment, we evaluate $10{,}000$ generated molecules using eight metrics: chemical validity, Fréchet ChemNet Distance (FCD)~\citep{preuer2018frechet}, the neighborhood subgraph pairwise distance kernel (NSPDK), novelty with respect to the training molecules, uniqueness (Unique), and structural similarity metrics, including scaffold similarity (Scaffold), fragment similarity (Fragment), and similarity to the nearest neighbor (SNN).

\subsection{Results} In \Cref{fig:acceleration_gen,tab:generation_quality}, we report the performance of generative models trained via Boltz2-based distillation. Representation alignment with Boltz2 accelerates training and yields the largest gains across most evaluation metrics. In contrast, alignment with existing foundation models, which are primarily pretrained for property prediction, yields marginal improvements. These results indicate that molecular representations vary in their effectiveness as supervision for generative modeling, and Boltz2, trained for protein-ligand co-folding, provides stronger structural supervision for molecular generation. Notably, none of the recent state-of-the-art graph diffusion models achieves dominant performance across validity, FCD, and NSPDK simultaneously, with each method excelling on only a subset of these metrics. Boltz2-based distillation, in contrast, improves all three, indicating that representation alignment yields coherent gains rather than trade-offs across metrics.

We further evaluate representation alignment on a larger generative model with an eight-layer Pairformer architecture. As demonstrated in \Cref{tab:generation_huge} in \Cref{appx:full_gen}, Boltz2-based distillation consistently improves generation quality, indicating that the benefit of Boltz2-based distillation is not limited to low-capacity generative models.

\begin{figure*}[t]
\centering
\includegraphics[width=0.95\linewidth]{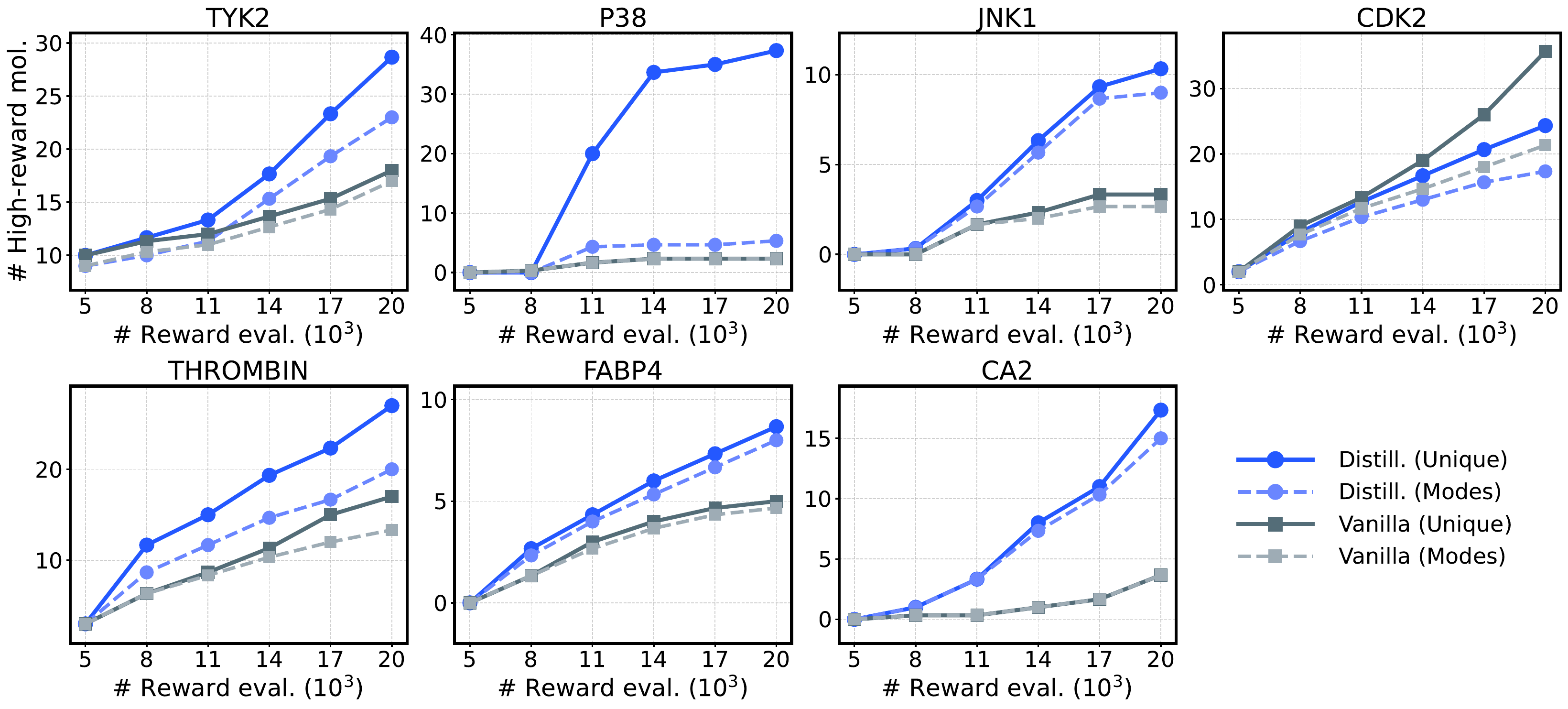}\quad\quad
\caption{\textbf{Results on structure-guided ligand discovery.} The results are averaged over three random seeds. Representation alignment with Boltz2 improves the sample efficiency for discovering high-score molecules that bind to target structures.}\label{fig:gflownet}
\vspace{-.1in}
\end{figure*}

\section{White-box Structure-guided Ligand Optimization}\label{sec:structure_guided}

We further evaluate Boltz2 representations in an online structure-guided ligand discovery setting \citep{passaro2025boltz2,cretu2025synflownet}. The task is formulated as online reinforcement learning, where a molecular generative policy is iteratively updated using predicted binding affinities of generated molecules for a target protein. The objective is to optimize generated molecules with respect to Boltz2 affinity as the reward signal. Here, we treat Boltz2 as a white-box reward teacher rather than only a black-box scalar oracle to better maximize the reward. For each generated molecule, Boltz2 produces both the scalar affinity reward and intermediate ligand representations. We use these representations as dense supervision for the policy, improving credit assignment during policy optimization. To our knowledge, this is the first use of representation alignment \citep{yu2024representation} to improve online reinforcement learning.



\subsection{Experimental Setup}

\textbf{Target proteins.} We use seven target proteins as benchmarks. 
First, we consider four benchmark targets from the Boltz2 paper, where Boltz2 shows strong binding affinity prediction performance \citep{passaro2025boltz2}: TYK2, CDK2, JNK1, and P38. As these targets are all kinases, we additionally include three proteins from distinct classes: CA2 as a metalloenzyme, THROMBIN as a serine protease, and FABP4 as a lipid-binding protein, covering diverse protein domains and ligand interaction mechanisms.

\textbf{Implementation details.} We use the SynFlowNet-Boltz pipeline introduced in the Boltz2 paper~\citep{passaro2025boltz2}. In this pipeline, the SynFlowNet policy, parameterized by a four-layer graph transformer~\cite{yun2019graph,cretu2025synflownet}, sequentially selects actions to complete a ligand molecule at each iteration. Then, Boltz2 computes binding affinity scores for the generated molecules based on protein-ligand structure predictions. These scores are used as rewards to update the policy toward generating higher-score ligands. 

In our experiments, we extend this setup by incorporating representation alignment-based distillation into the policy. Specifically, we maximize the cosine similarity between the policy’s second-layer single representations and corresponding Boltz2 single representations. Note that representations are aligned on ligand molecules, using Boltz2 ligand representations that are obtained as a byproduct of the Boltz2 binding affinity computation. The overall implementation and hyperparameters follow prior settings \citep{passaro2025boltz2}, with details provided in \Cref{supp:molopt_details}.

\textbf{Metrics.} We evaluate the performance using two metrics. First, we measure the number of high-score molecules discovered as a function of the number of reward evaluations during online training, reflecting sample efficiency. A molecule is considered high-reward if its Boltz2 screening score~\citep{passaro2025boltz2} exceeds $0.75$. For CA2, we adopt a higher threshold of $1.2$ to account for metal coordination effects. Second, we measure the number of modes, defined as the number of distinct high-scoring molecules with pairwise Tanimoto similarity below $0.6$.

\subsection{Results}

We present the results of structure-guided ligand discovery in \Cref{fig:gflownet}. For six target proteins, one can see that incorporating representation alignment with Boltz2 consistently increases the number of discovered high-reward molecules compared to the vanilla pipeline under the same reward evaluation budget, i.e., the same number of Boltz2 binding affinity computations for newly generated molecules. Note that representation alignment with Boltz2 also promotes exploration, as evidenced by an increase in the number of discovered high-reward modes.

Overall, these results show that exposing Boltz2’s intermediate ligand representations provides a stronger training signal than using its binding affinity rewards alone in online structure-guided discovery. By distilling interaction-aware representations computed during binding affinity prediction, the policy is guided toward high-reward regions, resulting in a faster discovery of high-affinity and diverse ligands. In \Cref{fig:distill_compare} of \Cref{supp:distill_compare}, we additionally conduct ablation studies on variants of representation alignment.

\section{Analyses on Boltz2 Representations}\label{sec:ablation}





\begin{figure}[t]
\centering
\vspace{-.08in}

\begin{minipage}[c]{0.45\linewidth}
\centering
\begin{subfigure}[t]{0.49\linewidth}
\centering
\includegraphics[width=\linewidth]{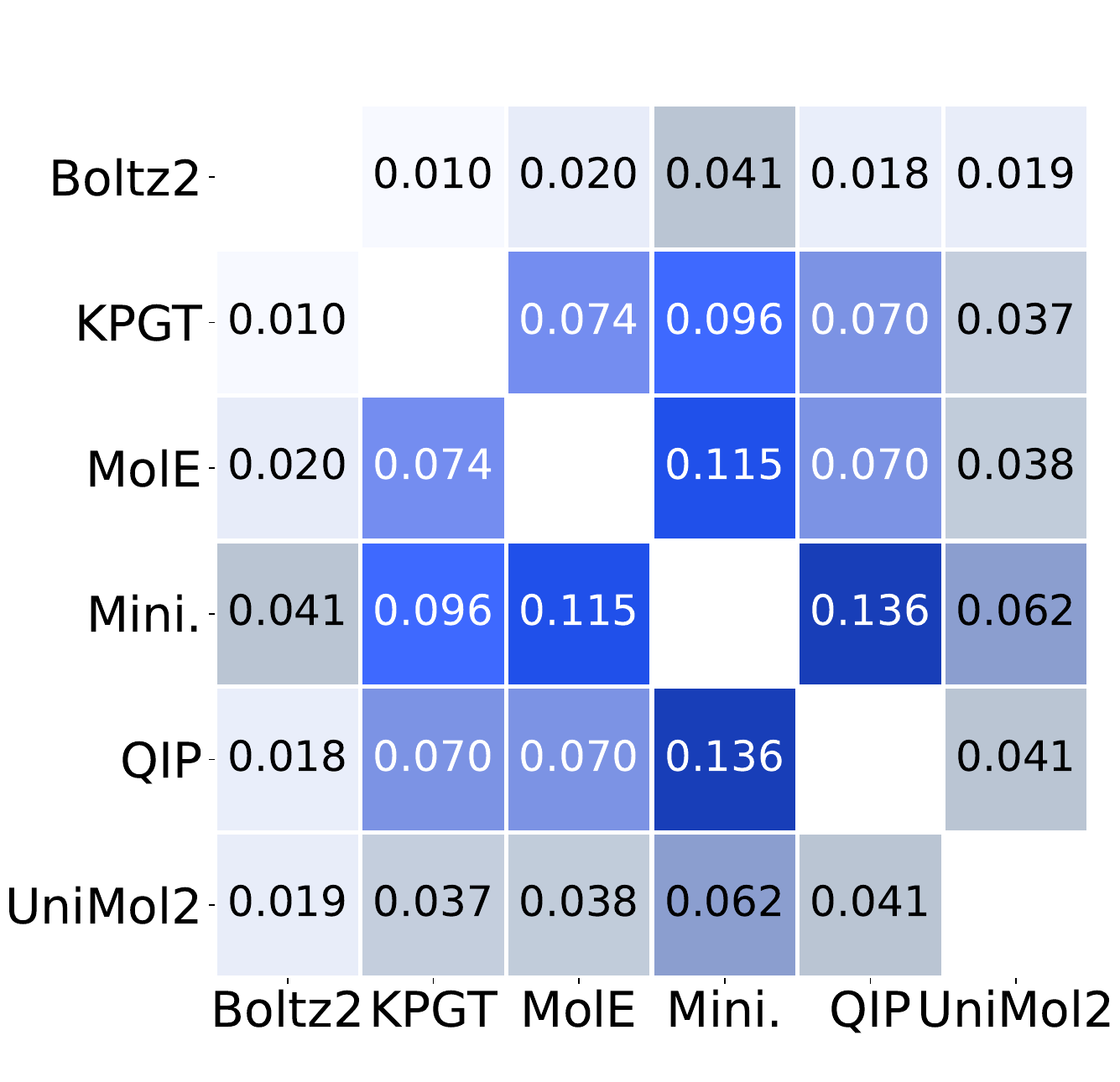}
\caption{CKNNA ($K=5$)}
\end{subfigure}
\begin{subfigure}[t]{0.49\linewidth}
\centering
\includegraphics[width=\linewidth]{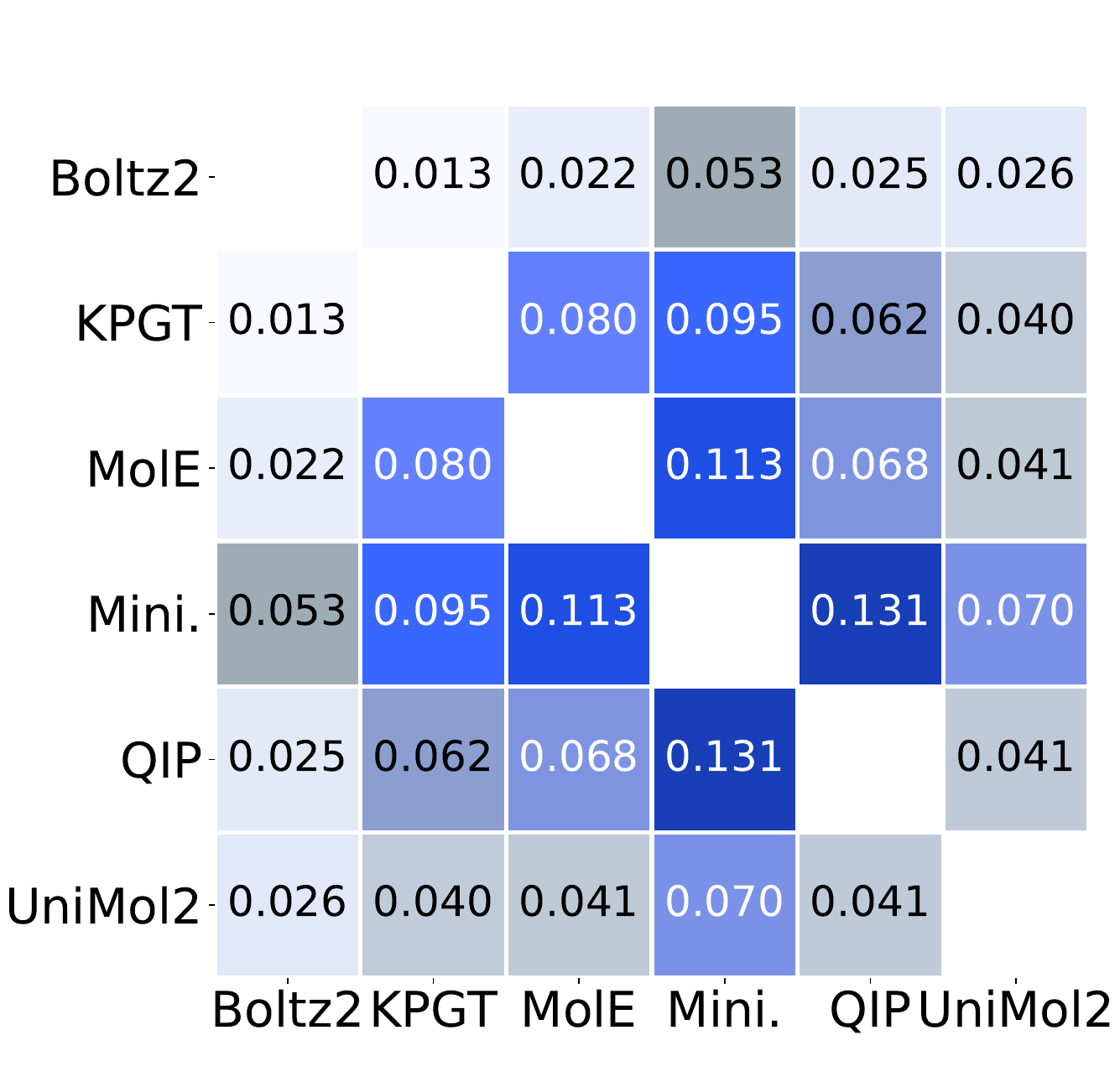}
\caption{CKNNA ($K=50$)}
\end{subfigure}
\end{minipage}
\;\;
\begin{minipage}[c]{0.50\linewidth}
\centering
\vspace{0.35cm}
\scalebox{0.84}{
\begin{tabular}{l c c c}
\toprule
\textbf{Model} & CYP2C9 V. & CYP2D6 V. & CYP3A4 V. \\
\midrule
\textbf{MolE}   & 0.80 & 0.68 & 0.87 \\ 
\textbf{Mini.}  & 0.82 & 0.72 & 0.88 \\
\midrule
\textbf{Boltz2} & 0.82 & 0.69 & 0.85 \\
\textbf{Boltz2}$^{\mathrm{MolE}}$  & \textbf{0.88{\scriptsize (+.06)}} & \textbf{0.77{\scriptsize (+.08)}} & \textbf{0.89{\scriptsize (+.04)}} \\
\textbf{Boltz2}$^{\mathrm{Mini.}}$ & 0.86{\scriptsize (+.04)} & 0.72{\scriptsize (+.03)} & 0.88{\scriptsize (+.03)} \\
\bottomrule
\end{tabular}
}
\end{minipage}

\caption{\textbf{Representation complementarity of Boltz2 with existing molecular foundation models.}
\textbf{Left:} Boltz2 exhibits relatively weak CKNNA with existing molecular foundation models.
\textbf{Right:} Combining Boltz2 with a less aligned representation (MolE) yields larger gains on some benchmarks than combining with a more aligned representation (MiniMol).}
\label{fig:cknna_ensemble}
\vspace{-.12in}
\end{figure}

\textbf{Boltz2 yields a distinct representation space that complements foundation models trained on standalone molecules.} We analyze representation alignment between Boltz2 and existing molecular foundation models using the CKNNA described in \Cref{sec:mol_gen}. According to the results in the left figure of \Cref{fig:cknna_ensemble}, Boltz2 shows relatively low average alignment with existing molecular foundation models trained on standalone molecular data. This suggests that Boltz2's co-folding learns a representation space that is distinct from those of existing foundation models, while consistently achieving strong performance across downstream tasks.

This observation raises the question of whether low representation alignment reflects complementary information that can be exploited through ensembling. While the performance improvements of Boltz2$^{\text{Mini.}}$ in \Cref{tab:admet_highlevel} support this hypothesis, we further test this using an ensemble with lower alignment to Boltz2, namely Boltz2+MolE. As shown in the right table of \Cref{fig:cknna_ensemble}, the Boltz2+MolE ensemble yields larger gains on several benchmarks than Boltz2+MiniMol, despite MiniMol exhibiting stronger performance than MolE. We provide full results in \Cref{tab:admet_boltz_only} of \Cref{supp:full_admet}.

\begin{wraptable}{r}{0.5\linewidth}
\vspace{-.13in}
\caption{\textbf{Boltz2 performance with protein-context.} Including protein context improves performance on enzyme inhibition tasks.}
\label{tab:metabolism_protein}
\vspace{-.05in}
\centering
\scalebox{0.86}{
\begin{tabular}{l c c }
\toprule
\textbf{Method} &
\textbf{Without protein} &
\textbf{With protein} \\
\hline
\rowcolor{gray!12}\multicolumn{3}{l}{\textit{Molecular (global) structure inhibits protein}} \\
CYP2C9 V. $\uparrow$  & 0.82 & 0.83{\scriptsize(+.01)} \\
CYP2D6 V. $\uparrow$  & 0.69 & 0.72{\scriptsize(+.03)} \\
CYP3A4 V. $\uparrow$  & 0.85 & 0.87{\scriptsize(+.02)} \\
\bottomrule
\end{tabular}}
\vspace{-.1in}
\end{wraptable}

\textbf{Incorporating protein context as inputs can improve downstream performance.} In \Cref{sec:admet}, we omit protein inputs to use isolated molecular representations. Here, we study how incorporating protein context affects downstream performance. We focus on metabolism tasks, which predict interactions between proteins and small molecules. We incorporate CYP2C9, CYP2D6, and CYP3A4 protein sequences and evaluate representations.

We report the results in \Cref{tab:metabolism_protein}. One can see that incorporating protein context improves performance on inhibition-related metabolism tasks. This indicates that Boltz2 representations can further benefit from additional task-relevant protein context. 

\begin{wraptable}{r}{0.5\linewidth}
\caption{\textbf{Layer-wise evaluation of Boltz2 representations.} \textbf{Bold} numbers indicate the best performance. Performance varies across layers depending on the task, while multi-layer concatenation generally yields promising results.}
\label{tab:layerwise}
\centering
\scalebox{0.86}{
\begin{tabular}{l c c c c c}
\toprule
\textbf{Method} &
\textbf{16th} &
\textbf{32nd} &
\textbf{48th} &
\textbf{64th} &
\textbf{Concat} \\
\midrule
\textbf{Solubility $\downarrow$}
& 0.67 & 0.67 & 0.66 & \textbf{0.64} & 0.66 \\
\textbf{BBB $\uparrow$}
& 0.92 & 0.92 & 0.92 & 0.91 & \textbf{0.93} \\
\textbf{CYP2C9 V. $\uparrow$}
& 0.81 & \textbf{0.82} & \textbf{0.82} & 0.81 & \textbf{0.82} \\
\textbf{Half Life $\uparrow$}
& 0.60 & \textbf{0.63} & 0.61 & 0.62 & 0.62 \\
\textbf{LD50 $\downarrow$}
& 0.41 & 0.41 & 0.40 & \textbf{0.39} & 0.40 \\
\bottomrule
\end{tabular}}
\vspace{-.1in}
\end{wraptable}

\textbf{Downstream performance varies across Boltz2 layers.} We analyze the contribution of representations extracted from different layers of Pairformer trunk to downstream performance. Specifically, we extract pair representations from the $\{16, 32, 48, 64\}$-th layers and evaluate them using one task from each ADMET category. We use the same probing setup as in \Cref{sec:admet}. As shown in \Cref{tab:layerwise}, performance varies across layers depending on the task, indicating that task-relevant information is distributed across the depth of the Pairformer trunk. While no single layer consistently dominates, concatenating representations from multiple layers yields strong overall performance.

\section{Conclusion}\label{sec:discussion}

In this work, we identify protein-ligand co-folding structures as a new source of supervision for atom-level representation learning. We show that Boltz2 representations transfer to small-molecule tasks and achieve strong performance on molecular property prediction, molecular generation, and structure-guided ligand discovery, while complementing representations from existing standalone molecular foundation models. Our results highlight co-folding-based pretraining as an effective strategy for small-molecule representation learning and position Boltz2 as a strong off-the-shelf baseline for small-molecule foundation models. An interesting avenue for future work is to investigate the use of protein context as a ``task-specific prompting strategy'' to reinforce small-molecule representations. In addition, we are the first to show that representation alignment-based distillation is effective for online reinforcement learning, which could be further explored in other general domains.



\textbf{Limitations.} First, while we position Boltz2 as an atom-level foundation model, prior work shows co-folding models lack complete atom-level understanding~\citep{brotzakis2025alphafold}. Our metabolism substrate results are consistent with this, as these require local atomistic mechanisms. However, we find that UniMol2 also exhibits the same weakness, suggesting a broader limitation of current 3D structure-based atom-level representation learning. By contrast, high-performing baselines on metabolism substrate instead rely on bioassay supervision that leaks label information~\citep{sypetkowski2024scalability,koleiev2026critical}, leaving genuine atom-level understanding without such supervision an open problem.  Second, we use the standard frozen-backbone setting through probing and distillation, and leave backbone fine-tuning for future work.



\bibliography{example_paper.bib}
\bibliographystyle{unsrt}

\newpage
\appendix
\crefname{section}{Appendix}{Appendices}
\Crefname{section}{Appendix}{Appendices}
\crefformat{section}{Appendix~#2{\color{blue(pigment)}#1}#3}




\newpage 

\section{Details in ADMET Benchmark}

\subsection{TDC ADMET Benchmark}\label[appendix]{supp:admet_statistics}

\begin{table}[h]
\centering
\vspace{-.1in}
\caption{\textbf{Data statistics and evaluation metrics of the TDC ADMET benchmark.} TDC ADMET benchmark is distributed through TDC, whose codebase is MIT-licensed.}
\label{tab:admet_stats_metrics}
\setlength{\tabcolsep}{10pt}
\begin{tabular}{l l l r}
\toprule
\textbf{Dataset} &
\textbf{Task} &
\textbf{Metric} &
\textbf{\# Molecules} \\
\midrule
Caco2 (Caco-2 Permeability)                & Regression     & MAE      & 906 \\
HIA (Human Intestinal Absorption)           & Classification & AUROC    & 578 \\
PgP (P-glycoprotein Inhibition)             & Classification & AUROC    & 1,213 \\
Bioavailability (Oral Bioavailability)      & Classification & AUROC    & 640 \\
Lipophilicity (LogD)                        & Regression     & MAE      & 4,200 \\
Solubility (Aqueous Solubility)             & Regression     & MAE      & 1,144 \\
BBB (Blood--Brain Barrier Penetration)      & Classification & AUROC    & 2,050 \\
PPBR (Plasma Protein Binding Rate)          & Regression     & MAE      & 1,797 \\
VDss (Volume of Distribution)              & Regression     & Spearman & 1,130 \\
CYP2C9 V.\ (CYP2C9 Inhibition)            & Classification & AUPRC    & 12,092 \\
CYP2D6 V.\ (CYP2D6 Inhibition)            & Classification & AUPRC    & 13,130 \\
CYP3A4 V.\ (CYP3A4 Inhibition)            & Classification & AUPRC    & 12,328 \\
CYP2C9 S.\ (CYP2C9 Substrate)             & Classification & AUPRC    & 666 \\
CYP2D6 S.\ (CYP2D6 Substrate)             & Classification & AUPRC    & 667 \\
CYP3A4 S.\ (CYP3A4 Substrate)             & Classification & AUROC    & 667 \\
Half Life                                   & Regression     & Spearman & 1,039 \\
Clearance H.\ (Hepatocyte)                 & Regression     & Spearman & 1,102 \\
Clearance M.\ (Microsome)                  & Regression     & Spearman & 1,102 \\
LD50 (Acute Toxicity)                       & Regression     & MAE      & 7,385 \\
hERG (hERG Blockade)                        & Classification & AUROC    & 648 \\
AMES (Mutagenicity)                         & Classification & AUROC    & 6,512 \\
DILI (Drug-Induced Liver Injury)            & Classification & AUROC    & 475 \\
\bottomrule
\end{tabular}
\vspace{-.1in}
\end{table}

\subsection{Implementation Details}\label[appendix]{supp:admet_details}

\textbf{Representation extraction.} In this section, we describe the pooling strategy for ADMET property prediction. Note that Boltz does not include a pooling layer and is not trained on pooled representations to extract informative fixed-length representations for downstream tasks, unlike existing small-molecule foundation models. A naive choice would be simple global mean pooling over all atom-pair entries, but this risks ignoring local features that are critical for tasks such as substrate prediction. We therefore apply a hybrid pooling strategy to capture diverse structural statistics of pair representations. Note that we apply the identical pooling strategy to UniMol2 for a fair comparison.

For Boltz2 and UniMol2, we concatenate pair representations from the $\{16, 32, 48, 64\}$-th layers, corresponding to the $1/4$, $2/4$, $3/4$, and final layers of the $64$-layer Pairformer. We then concatenate pooled representations over diagonal entries, bonded atom-pair entries, and all entries in the pair representation matrix. Specifically, we apply statistic pooling~\citep{okabe2018attentive}. 

Importantly, even with reduced pooling, Boltz representations outperform existing molecular representation models (\Cref{supp:pooling_ablation}). However, our hybrid pooling does not recover performance on substrate prediction tasks, suggesting that the substrate gap is unlikely to be a pooling artifact and instead reflects a more fundamental limitation of structure-based representations on local atomistic mechanisms (see limitations in \Cref{sec:discussion}).



\textbf{Probing network.} The probing network is implemented as a multi-layer perceptron (MLP), following prior works~\citep{klaser2024minimol,sypetkowski2024scalability}. The hidden dimension and the number of layers of the MLP are selected from $\{512, 1024, 2048\}$ and $\{2,3,4\}$, respectively. The hidden representation at each layer is concatenated with the input, following the model design in prior work~\citep{klaser2024minimol}. The dropout rate is fixed to $0.0$. The learning rate is selected from $\{1\mathrm{e}{-4}, 3\mathrm{e}{-4}, 5\mathrm{e}{-4}\}$, and the weight decay is $1\mathrm{e}{-5}$. Models are trained for $25$ or $200$ epochs. We use batch size $32$. Early stopping is applied with a patience of $25$ epochs, and a cosine learning-rate scheduler is used. Hyperparameters are selected based on validation performance. 

\newpage

\section{Full results of Boltz on TDC ADMET Benchmarks}\label[appendix]{supp:full_admet}

\begin{table}[h]
\centering
\caption{\textbf{Full results of representation ensembling on the ADMET benchmark.}}
\label{tab:admet_boltz_only}
\setlength{\tabcolsep}{8pt}
\begin{tabular}{l c c c}
\toprule
\textbf{Dataset} &
\textbf{Boltz2} &
\textbf{Boltz2$^{\mathrm{Mini.}}$} &
\textbf{Boltz2$^{\mathrm{MolE}}$} \\
\hline
\rowcolor{gray!12}\multicolumn{4}{l}{\textit{Absorption}} \\
Caco2 $\downarrow$              & 0.30 $\pm$ .004 & 0.30 $\pm$ .009 & 0.30 $\pm$ .005 \\
HIA $\uparrow$                  & 0.99 $\pm$ .001 & 0.99 $\pm$ .001 & 0.99 $\pm$ .001 \\
Pgp $\uparrow$                  & 0.93 $\pm$ .002 & 0.93 $\pm$ .002 & 0.93 $\pm$ .003 \\
Bioavailability $\uparrow$      & 0.75 $\pm$ .009 & 0.77 $\pm$ .006 & 0.72 $\pm$ .009 \\
Lipophilicity $\downarrow$      & 0.45 $\pm$ .005 & 0.41 $\pm$ .003 & 0.43 $\pm$ .007 \\
Solubility $\downarrow$         & 0.66 $\pm$ .013 & 0.64 $\pm$ .017 & 0.65 $\pm$ .003 \\
\hline
\rowcolor{gray!12}\multicolumn{4}{l}{\textit{Distribution}} \\
BBB $\uparrow$                  & 0.93 $\pm$ .003 & 0.94 $\pm$ .001 & 0.92 $\pm$ .005 \\
PPBR $\downarrow$               & 7.65 $\pm$ .063 & 7.59 $\pm$ .090 & 7.34 $\pm$ .074 \\
VDss $\uparrow$                 & 0.74 $\pm$ .010 & 0.75 $\pm$ .007 & 0.75 $\pm$ .008 \\
\hline
\rowcolor{gray!12}\multicolumn{4}{l}{\textit{Metabolism}} \\
CYP2C9 V.\ $\uparrow$          & 0.82 $\pm$ .003 & 0.86 $\pm$ .003 & 0.88 $\pm$ .002 \\
CYP2D6 V.\ $\uparrow$          & 0.69 $\pm$ .002 & 0.72 $\pm$ .004 & 0.77 $\pm$ .001 \\
CYP3A4 V.\ $\uparrow$          & 0.85 $\pm$ .001 & 0.88 $\pm$ .002 & 0.89 $\pm$ .002 \\
CYP2C9 S.\ $\uparrow$          & 0.36 $\pm$ .015 & 0.36 $\pm$ .029 & 0.37 $\pm$ .021 \\
CYP2D6 S.\ $\uparrow$          & 0.52 $\pm$ .012 & 0.51 $\pm$ .021 & 0.55 $\pm$ .024 \\
CYP3A4 S.\ $\uparrow$          & 0.62 $\pm$ .011 & 0.60 $\pm$ .012 & 0.64 $\pm$ .013 \\
\hline
\rowcolor{gray!12}\multicolumn{4}{l}{\textit{Excretion}} \\
Half Life $\uparrow$            & 0.62 $\pm$ .010 & 0.65 $\pm$ .015 & 0.66 $\pm$ .030 \\
Clearance H.\ $\uparrow$       & 0.62 $\pm$ .009 & 0.60 $\pm$ .018 & 0.57 $\pm$ .013 \\
Clearance M.\ $\uparrow$       & 0.61 $\pm$ .008 & 0.65 $\pm$ .011 & 0.65 $\pm$ .008 \\
\hline
\rowcolor{gray!12}\multicolumn{4}{l}{\textit{Toxicity}} \\
LD50 $\downarrow$               & 0.40 $\pm$ .004 & 0.40 $\pm$ .006 & 0.40 $\pm$ .003 \\
hERG $\uparrow$                 & 0.86 $\pm$ .009 & 0.86 $\pm$ .008 & 0.83 $\pm$ .008 \\
AMES $\uparrow$                 & 0.91 $\pm$ .002 & 0.91 $\pm$ .003 & 0.92 $\pm$ .001 \\
DILI $\uparrow$                 & 0.89 $\pm$ .008 & 0.87 $\pm$ .009 & 0.87 $\pm$ .009 \\
\bottomrule
\end{tabular}
\end{table}

We report the full results of representation ensembling in Supplementary Table~\ref{tab:admet_boltz_only}. Boltz$^{\text{MolE}}$ outperforms Boltz$^{\text{Mini.}}$ on some benchmarks, although MiniMol shows stronger standalone ADMET prediction performance than MolE.

\newpage

\section{Comparison with MolGPS at matched parameter scale}
\label[appendix]{app:scale}

\begin{table}[h]
\centering
\caption{\textbf{Boltz2 vs.\ MolGPS backbones at matched 1B parameter scale.}
We use numbers of three 1B MolGPS backbones (MPNN++, Transformer, GPS++) \citep{sypetkowski2024scalability}. \textbf{Bold} indicates the best performance.}
\label{tab:boltz_vs_molgps_1b}
\begin{tabular}{lcccc}
\toprule
\textbf{Dataset} & \textbf{MPNN++} & \textbf{Transformer} & \textbf{GPS++} & \textbf{Boltz2} \\
\hline
\rowcolor{gray!12}\multicolumn{5}{l}{\textit{Absorption}} \\
Caco2 $\downarrow$         & 0.58 & 0.36 & 0.40 & \textbf{0.30} \\
HIA $\uparrow$             & 0.96 & 0.93 & 0.96 & \textbf{0.99} \\
Pgp $\uparrow$             & \textbf{0.94} & 0.92 & \textbf{0.94} & 0.93 \\
Bioavailability $\uparrow$ & 0.70 & 0.70 & 0.69 & \textbf{0.75} \\
Lipophilicity $\downarrow$ & 0.55 & 0.52 & 0.49 & \textbf{0.45} \\
Solubility $\downarrow$    & 0.86 & 0.75 & 0.78 & \textbf{0.66} \\
\hline
\rowcolor{gray!12}\multicolumn{5}{l}{\textit{Distribution}} \\
BBB $\uparrow$    & 0.90 & 0.91 & 0.90 & \textbf{0.93} \\
PPBR $\downarrow$ & 9.09 & 13.36 & 10.91 & \textbf{7.65} \\
VDss $\uparrow$   & 0.60 & 0.65 & 0.65 & \textbf{0.74} \\
\hline
\rowcolor{gray!12}\multicolumn{5}{l}{\textit{Metabolism}} \\
CYP2C9 V.\ $\uparrow$ & \textbf{0.85} & {0.82} & {0.82} & {0.82} \\
CYP2D6 V.\ $\uparrow$ & 0.70 & \textbf{0.71} & 0.69 & 0.69 \\
CYP3A4 V.\ $\uparrow$ & \textbf{0.88} & \textbf{0.88} & \textbf{0.88} & 0.85 \\
CYP2C9 S.\ $\uparrow$ & 0.37 & \textbf{0.41} & 0.38 & 0.36 \\
CYP2D6 S.\ $\uparrow$ & \textbf{0.68} & 0.66 & 0.66 & 0.52 \\
CYP3A4 S.\ $\uparrow$ & {0.65} & \textbf{0.72} & 0.69 & 0.62 \\
\hline
\rowcolor{gray!12}\multicolumn{5}{l}{\textit{Excretion}} \\
Half Life $\uparrow$     & 0.44 & 0.56 & 0.53 & \textbf{0.62} \\
Clearance H.\ $\uparrow$ & 0.39 & 0.31 & 0.36 & \textbf{0.62} \\
Clearance M.\ $\uparrow$ & 0.61 & 0.50 & \textbf{0.62} & 0.61 \\
\hline
\rowcolor{gray!12}\multicolumn{5}{l}{\textit{Toxicity}} \\
LD50 $\downarrow$ & 0.72 & 0.67 & 0.69 & \textbf{0.40} \\
hERG $\uparrow$   & 0.83 & 0.81 & 0.79 & \textbf{0.86} \\
AMES $\uparrow$   & 0.81 & 0.83 & 0.83 & \textbf{0.91} \\
DILI $\uparrow$   & 0.85 & \textbf{0.90} & \textbf{0.90} & {0.89} \\
\bottomrule
\end{tabular}
\end{table}

\newpage

\section{Details in Molecular Generation Benchmark}\label[appendix]{supp:mol_gen_details}

\subsection{CKNNA Metric}\label[appendix]{supp:cknna}

We measure alignment using Centered Kernel Nearest-Neighbor Alignment metric (CKNNA)~\citep{huh2024platonic}, which evaluates local alignment between two representation spaces based on shared nearest-neighbor structure. Given a set of molecules $\{m_i\}_{i=1}^N$ and two representation models $f$ and $g$, yielding representations $\{f(m_i)\}_{i=1}^N$ and $\{g(m_i)\}_{i=1}^N$, CKNNA computes the alignment score as follows:
\begin{align*}
&\mathrm{ALIGN}\big(\{f(m_i)\}_{i=1}^N, \{g(m_i)\}_{i=1}^N\big) = \\
&\frac{1}{(N-1)^2}
\sum_{i,j}
\alpha(i,j)
\big(\langle f(m_i), f(m_j) \rangle - \mathbb{E}_l[\langle f(m_i), f(m_l) \rangle]\big)
\big(\langle g(m_i), g(m_j) \rangle - \mathbb{E}_l[\langle g(m_i), g(m_l) \rangle]\big),
\end{align*}
where $\alpha(i,j)$ selects pairs of samples that lie within local neighborhoods:
\begin{equation}
\alpha(i,j; k) =
\mathbbm{1}\big[
i \neq j \;\wedge\;
f(m_j) \in \mathrm{KNN}(f(m_i); k) \;\wedge\;
g(m_j) \in \mathrm{KNN}(g(m_i); k)
\big],
\end{equation}
and $\mathrm{KNN}(\cdot; k)$ denotes the set of $k$ nearest neighbors. CKNNA is then computed as
\begin{equation}
\mathrm{CKNNA} =
\frac{
\mathrm{ALIGN}\big(\{f(m_i)\}_{i=1}^N, \{g(m_i)\}_{i=1}^N\big)
}{
\sqrt{
\mathrm{ALIGN}\big(\{f(m_i)\}_{i=1}^N, \{f(m_i)\}_{i=1}^N\big)
\mathrm{ALIGN}\big(\{g(m_i)\}_{i=1}^N, \{g(m_i)\}_{i=1}^N\big)
}
}.
\end{equation}

\subsection{Implementation Details}\label[appendix]{supp:molgen_details}

\textbf{Denoising model.} The denoising network is parameterized by a four-layer Pairformer architecture with $4$ attention heads, using hidden dimensions of $256$ for single-node representations and $128$ for pair representations. Single-node representations are used to denoise atom attributes of the molecular graph using a two-layer MLP with a hidden dimension of $256$, while pair representations are used to denoise bond attributes using a two-layer MLP with a hidden dimension of $128$. Before being passed into the Pairformer, pair representations are initialized by transforming edge features, concatenations of atom-pair features, and a time conditioning signal with two-layer MLPs. Single representations are initialized from atom features and the time conditioning signal.

\textbf{Training configuration.} The learning rate is set to $3\mathrm{e}{-4}$ with a cosine learning-rate scheduler. The batch size is $512$, and models are trained for $100$ epochs. All other hyper-parameters, including weight decay, scaling, and noise scheduling, follow the default GruM configuration \citep{jo2023graph}.

\textbf{Representation alignment.} We adopt a representation alignment-based distillation objective to train molecular generative models, specifically GruM, with auxiliary supervision from Boltz2 representations. Given a noisy molecule $m_t$ and its corresponding clean molecule $m$, we define the training objective as follows:
\begin{equation*}
\mathcal{L}(m_t, m)
=
\mathcal{L}_{\text{GruM}}\big(s_{\theta}(m_t), m\big)
-
\lambda\cdot \cos\!\big(h_{\theta}(m_t), f(m)\big),
\end{equation*}
where $\mathcal{L}_{\text{GruM}}$ denotes the denoising loss of GruM, $s_{\theta}(m_t)$ is the denoising prediction, $h_{\theta}(m_t)$ denotes the output of the distillation network given the hidden representation of the generative model, and $f(m)$ denotes the frozen Boltz2 representation of the clean molecule. For distillation, we introduce a lightweight distillation network, a two-layer MLP with a hidden dimension of $1536$. The distillation network maps generative model representations to the Boltz2 representation space. Boltz2 representations $f(m)$ are precomputed for all molecules. We also flatten the single and pair representations when applying representation alignment, while masking out-of-range indices.

To be more specific, the cosine similarity is computed between the generative model representations and the Boltz2 representations, and $\lambda$ denotes the coefficient for representation alignment. We set $\lambda = 4$ in our experiments. Note that representation alignment is applied at the second Pairformer layer of the denoising model for both single and pair representations, with the alignment target defined as the concatenated representations from the $\{16,32,48,64\}$-th layers of the Boltz2 Pairformer trunk.

\newpage

\section{Additional results on unconditional graph generation}\label[appendix]{appx:full_gen}

\newcolumntype{L}[1]{>{\raggedright\arraybackslash}p{#1}}
\newcolumntype{C}[1]{>{\centering\arraybackslash}p{#1}}

\begin{table*}[h]
\caption{\textbf{Full results on unconditional molecular generation.} GruM$^{\ast}$ denotes GruM parameterized with a Pairformer. Mean $\pm$ standard deviation over three random seeds is reported; entries with $\pm\textsc{x.xx}$ denote that the standard deviation is not reported in the original paper. \textbf{Bold} numbers indicate the best performance. Representation alignment with \hlblue{Boltz2} accelerates the training of molecular generative models to produce higher-quality molecules compared to the baselines.}
\label{tab:full_gen}
\centering
\scalebox{0.675}{
\begin{tabular}{l c c c c c c c c }
\toprule
{\pz\pz\textbf{Method}} &
{\textbf{Valid $\uparrow$}} &
{\textbf{FCD $\downarrow$}} &
{\textbf{NSPDK $\downarrow$}} &
{\textbf{Novel $\uparrow$}} &
{\textbf{Unique $\uparrow$}} &
{\textbf{Scaffold $\uparrow$}} &
{\textbf{Fragment $\uparrow$}} &
{\textbf{SNN $\uparrow$}} \\
\hline
\rowcolor{gray!12}\multicolumn{9}{l}{\textit{State-of-the-art diffusion-based graph generative models}} \\
\textbf{GruM}    & $98.65_{\pm 0.25}$ & $\pz2.26_{\pm 0.08}$ & $0.0015_{\pm 0.0003}$ & $99.98_{\pm 0.02}$ & $99.97_{\pm 0.03}$ & $0.5299_{\pm 0.0441}$ & -- & -- \\
\textbf{GBD}     & $97.87_{\pm \textsc{x.xx}}$ & $\pz2.25_{\pm \textsc{x.xx}}$ & $0.0018_{\pm \textsc{x.xxxx}}$ & -- & -- & $0.5042_{\pm \textsc{x.xxxx}}$ & -- & -- \\
\textbf{DeFoG}   & $99.22_{\pm 0.08}$ & $\pz1.42_{\pm 0.02}$ & $0.0008_{\pm 0.0001}$ & -- & $99.99_{\pm 0.01}$ & $\textbf{0.5903}_{\pm 0.0099}$ & -- & -- \\
\textbf{TopBF}   & $99.37_{\pm \textsc{x.xx}}$ & $\pz1.39_{\pm \textsc{x.xx}}$ & $0.0008_{\pm \textsc{x.xxxx}}$ & -- & -- & $0.5372_{\pm \textsc{x.xxxx}}$ & -- & -- \\
\textbf{Marg. SID}     & $99.50_{\pm \textsc{x.xx}}$ & $\pz2.01_{\pm \textsc{x.xx}}$ & $0.0021_{\pm \textsc{x.xxxx}}$ & -- & -- & -- & -- & -- \\
{\textbf{GruM}$^{\ast}$}
& $99.36_{\pm 0.09}$ & $\pz1.50_{\pm 0.07}$ & $0.0007_{\pm 0.0001}$ & $99.99_{\pm 0.01}$ & $\textbf{100.00}_{\pm 0.00}$ & $0.4923_{\pm 0.0074}$ & $0.9852_{\pm 0.0017}$ & $0.3697_{\pm 0.0028}$ \\
\hline
\rowcolor{gray!12}\multicolumn{9}{l}{\textit{Applying representation alignment-based distillation to GruM$^{\ast}$}} \\
\textbf{+UniMol2}
& $99.60_{\pm 0.07}$ & $\pz1.42_{\pm 0.03}$ & $0.0006_{\pm 0.0000}$ & $\textbf{100.00}_{\pm 0.00}$ & $\textbf{100.00}_{\pm 0.00}$ & $0.4864_{\pm 0.0133}$ & $0.9867_{\pm 0.0007}$ & $0.3721_{\pm 0.0002}$ \\

\textbf{+MolE}
& $99.55_{\pm 0.02}$ & $\pz1.40_{\pm 0.01}$ & $0.0006_{\pm 0.0000}$ & $\textbf{100.00}_{\pm 0.00}$ & $\textbf{100.00}_{\pm 0.00}$ & $0.4932_{\pm 0.0041}$ & $0.9864_{\pm 0.0004}$ & $0.3737_{\pm 0.0007}$ \\

\textbf{+KPGT}
& $99.37_{\pm 0.05}$ & $\pz1.46_{\pm 0.05}$ & $0.0006_{\pm 0.0000}$ & $99.99_{\pm 0.01}$ & $99.99_{\pm 0.01}$ & $0.4800_{\pm 0.0299}$ & $0.9860_{\pm 0.0009}$ & $0.3702_{\pm 0.0020}$ \\

\textbf{+Mini.}
& $99.48_{\pm 0.09}$ & $\pz1.45_{\pm 0.04}$ & $0.0007_{\pm 0.0001}$ & $99.99_{\pm 0.01}$ & $\textbf{100.00}_{\pm 0.00}$ & $0.4812_{\pm 0.0234}$ & $0.9856_{\pm 0.0009}$ & $0.3710_{\pm 0.0023}$ \\

\textbf{+QIP}
& $99.37_{\pm 0.09}$ & $\pz1.43_{\pm 0.05}$ & $0.0007_{\pm 0.0001}$ & $\textbf{100.00}_{\pm 0.00}$ & $\textbf{100.00}_{\pm 0.00}$ & $0.5242_{\pm 0.0071}$ & $0.9864_{\pm 0.0003}$ & $0.3718_{\pm 0.0015}$ \\

\rowcolor{aliceblue!60}
\textbf{+Boltz2}
& $\textbf{99.65}_{\pm 0.11}$ & $\pz\textbf{1.31}_{\pm 0.01}$ & $\textbf{0.0005}_{\pm 0.0000}$ & $\textbf{100.00}_{\pm 0.00}$ & $\textbf{100.00}_{\pm 0.00}$ & $0.5064_{\pm 0.0187}$ & $\textbf{0.9881}_{\pm 0.0003}$ & $\textbf{0.3766}_{\pm 0.0014}$ \\
\bottomrule
\end{tabular}}
\vspace{-.1in}
\end{table*}

\begin{table}[h]
\caption{\textbf{Distillation into a large-scale model.} Boltz2 representation improves molecular generation performance on a large-scale backbone.}
\label{tab:generation_huge}
\centering
\begin{tabular}{l c c c}
\toprule
\textbf{Method} &
\textbf{Validity $\uparrow$} &
\textbf{FCD $\downarrow$} &
\textbf{NSPDK $\downarrow$} \\
\midrule
\textbf{PairFormer}
& 99.41 & 1.49 & 0.0007 \\ 
\textbf{Align. w/ Boltz2} & 99.65 & 1.31 & 0.0005 \\ 
\midrule
\textbf{Pairformer}$^{\text{large}}$
& 99.56 & 1.35 & 0.0006 \\ 
\textbf{w/ Boltz2}
& \textbf{99.76} & \textbf{1.28} & \textbf{0.0005} \\ 
\bottomrule
\end{tabular}
\end{table}

\newpage

\section{Details in SynFlowNet-Boltz}\label[appendix]{supp:synflownet_details}

\begin{figure}[t]
\centering
\centering \includegraphics[width=0.8\linewidth]{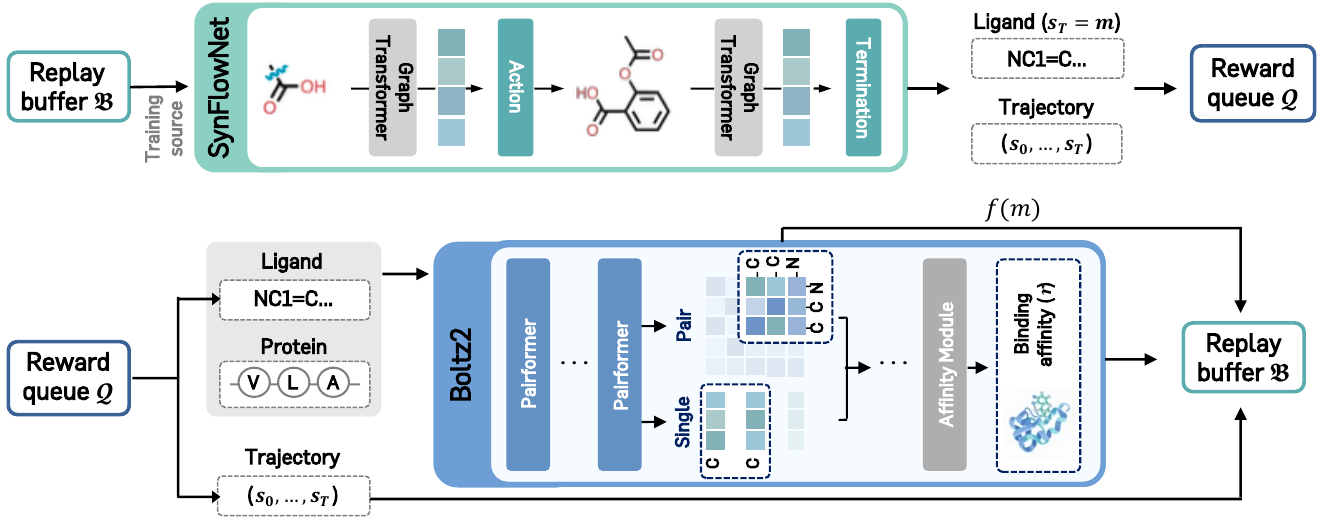}
\caption{\textbf{Online SynFlowNet-Boltz ligand discovery pipeline.} The top and bottom panels illustrate the processes of molecular generation with SynFlowNet and evaluation with Boltz2, respectively. Both  processes are asynchronous.}\label{fig:synflownet_pipeline}
\end{figure}

\subsection{Implementation Details}\label[appendix]{supp:molopt_details}

We use SynFlowNet-Boltz ligand discovery pipeline introduced in prior work \citep{passaro2025boltz2,cretu2025synflownet}. The task is formulated as an online reinforcement learning problem, where a molecular policy is iteratively updated based on binding affinity rewards computed by Boltz2. In this experiment, we use the official codebase of SynFlowNet-Boltz.

\textbf{SynFlowNet.} We use SynFlowNet~\citep{cretu2025synflownet}, a widely studied GFlowNet framework~\citep{bengio2021flow,malkin2022trajectory,bengio2023gflownet,kimlocal,janglearning,seogenerative}, for synthesizable molecular generation, which defines a reaction-based action space.

\textbf{Policy network.} At each generation step, the policy $P_F$ sequentially selects reaction templates and reactants to construct a molecule, yielding a trajectory $\tau = (s_0, \ldots, s_T)$ of molecular construction steps. The policy is parameterized by a four-layer graph transformer \citep{yun2019graph}, and we do not modify the original implementation \citep{passaro2025boltz2,cretu2025synflownet}. The graph transformer outputs $128$-dimensional atom-wise single representations, which are pooled and passed through an MLP to produce a distribution over valid actions. The action space is the same as that of SynFlowNet~\citep{cretu2025synflownet}, including reactant and reaction template selections. The backward policy in SynFlowNet is also implemented as a four-layer graph transformer and trained using a maximum likelihood–based approach, following prior implementations~\citep{cretu2025synflownet,jang2024pessimistic}.

\textbf{Reward computation.} For each generated terminal molecule $s_T=m$, Boltz2 computes a reward $r$ (a screening score) as follows:
\begin{equation*}
r=\text{max}\big(\frac{-\text{affinity}+2}{4},0\big)\cdot\text{likelihood}
\end{equation*}
where $\text{affinity}$ is the Boltz2-predicted binding affinity score for molecule $m$, and $\text{likelihood}$ denotes the Boltz2-predicted likelihood of binding to the corresponding binding site of the target protein. 

Here, Boltz2 uses single and pair representations, along with 3D structure prediction, for protein–ligand binding affinity predictions. Thus, we can naturally obtain single representations of ligand molecules within the SynFlowNet–Boltz pipeline.

\textbf{Representation alignment.} In addition to scalar rewards, we incorporate representation alignment to provide auxiliary supervision for policy training.
Specifically, for each terminal-state molecule, we extract single representations from Boltz2 that are computed during the binding affinity prediction process. We align these representations with the hidden single representations from the second layer of the policy network by maximizing cosine similarity. Here, we basically consider representation alignment to terminal-state molecules $s_T=m$:
\begin{equation*}
\mathcal{L}_{\text{align.}}(m, f(m))=
-\lambda\cdot\cos\!\big(h_{\theta}(m), f(m)\big),
\end{equation*}
where $h_{\theta}(m)$ denotes the output of the distillation network given the hidden representation of the policy network, and $f(m)$ denotes the frozen Boltz2 representation. For distillation, we also introduce a lightweight distillation network parameterized as a two-layer MLP with a hidden dimension of $512$. We set the representation alignment coefficient to $\lambda = 10$. 

To be specific, representation alignment is applied at the second graph transformer layer of the policy model, aligning its representations with those obtained from the Boltz2 Pairformer trunk. Since the graph transformer in the policy network outputs only single atom-wise representations, alignment is enforced only for single representations, and Boltz2 pair representations are omitted. In this setting, Boltz2 representations span both ligand atoms and protein residues. We therefore slice the representations to retain only the ligand atom representations $f(m)$.


As an ablation study, we also study intermediate-state distillation as an ablation in \Cref{supp:distill_compare}. This is defined as follows:
\begin{equation*}
\mathcal{L}^{\text{inter.}}_{\text{align.}}(\tau=(s_0,\ldots,s_T),f(m))=
-\lambda\cdot\sum^{T}_{t=0}\cos\!\big(
h_{\theta}(s_t),
f(m)_{\mathcal{I}(s_t)}
\big).
\end{equation*}
Here, $\mathcal{I}(s_t)$ denotes the index set of atoms in the terminal molecule $s_T$ that correspond to the molecular substructure present at the intermediate state $s_t$, and $f(m)_{\mathcal{I}(s_t)}$ denotes the representations of this corresponding substructure. In this ablation objective, an intermediate molecule $s_t$ may lead to multiple terminal molecules through different trajectory actions, resulting in multiple possible alignment targets for $h_{\theta}(s_t)$. To address this, we restrict alignment to representations of terminal molecules in the top $10\%$ by reward.

\begin{algorithm}[t]
\caption{Learning SynFlowNet with Representation Alignment}
\label{alg:synflownet}
\begin{algorithmic}[1]
   \STATE Initialize replay buffer $\mathcal{B}$, trajectory queue $\mathcal{Q}$, and SynFlowNet policy $P_{\theta}$
   \REPEAT
   \STATE Sample a batch of molecular generative trajectories $\{\tau^{(k)}=(s^{(k)}_0,\ldots,s^{(k)}_T=m^{(k)})\}_{k=1}^{K}$ from the policy $P_{\theta}$
   \STATE Update $\mathcal{Q} \leftarrow \mathcal{Q} \cup \{\tau^{(k)}\}_{k=1}^{K}$
   \STATE Sample a batch of trajectories with rewards and molecular representations $\{\tau^{(k)}, r^{(k)}, f(m^{(k)})\}_{k=1}^{K}$ from $\mathcal{B}$
   \STATE Update $P_F$ by minimizing $\mathcal{L}_{\text{SynFlowNet}}(\tau,r)+ \mathcal{L}_{\text{align.}}(m,f(m))$ using $\{\tau^{(k)}, r^{(k)}, f(m^{(k)})\}_{k=1}^{K}$
   \UNTIL{converged}
\end{algorithmic}
\end{algorithm}

\begin{algorithm}[t]
\caption{Asynchronous Boltz2 Worker}
\label{alg:boltz_worker}
\begin{algorithmic}[1]
   \REPEAT
   \STATE Sample the latest trajectories $\{\tau^{(k)}\}_{k=1}^{K}$ from the queue $\mathcal{Q}$
   \STATE Compute rewards and molecular representations $\{(r^{(k)}, f(m^{(k)}))\}_{k=1}^{K}$ using Boltz2
   \STATE Update $\mathcal{B} \leftarrow \mathcal{B} \cup \{\tau^{(k)}, r^{(k)}, f(m^{(k)})\}_{k=1}^{K}$
   \UNTIL{converged}
\end{algorithmic}
\end{algorithm}

\textbf{Optimization and training.} The policy is trained using SynFlowNet objective $\mathcal{L}_{\text{SynFlowNet}}$, augmented with the representation alignment loss $\mathcal{L}_{\text{align.}}$. As illustrated in \Cref{fig:synflownet_pipeline}, \Cref{alg:synflownet}, and \Cref{alg:boltz_worker}, policy optimization and reward computation of SynFlowNet-Boltz are decoupled and executed asynchronously \citep{passaro2025boltz2}. The policy process continuously samples trajectories, pushes them to the reward queue $\mathcal{Q}$, and trains on trajectories, rewards, and representations sampled from the replay buffer $\mathcal{B}$. The Boltz2 worker process computes binding affinity rewards and representations for queued trajectories and pushes them to the replay buffer $\mathcal{B}$. We use four NVIDIA B200 GPUs for policy training and run $16$ parallel Boltz2 worker processes for asynchronous reward computation.

Note that we initialize the replay buffer $\mathcal{B}$ with $5{,}000$ trajectories sampled from multiple warm-up policies. All other implementation details, including replay buffer size, reward normalization, and exploration strategy, follow prior work without modification \citep{passaro2025boltz2}.

\subsection{Intermediate State Distillation}\label[appendix]{supp:distill_compare}

\begin{figure}[t]
\centering
\centering \includegraphics[width=0.5\linewidth]{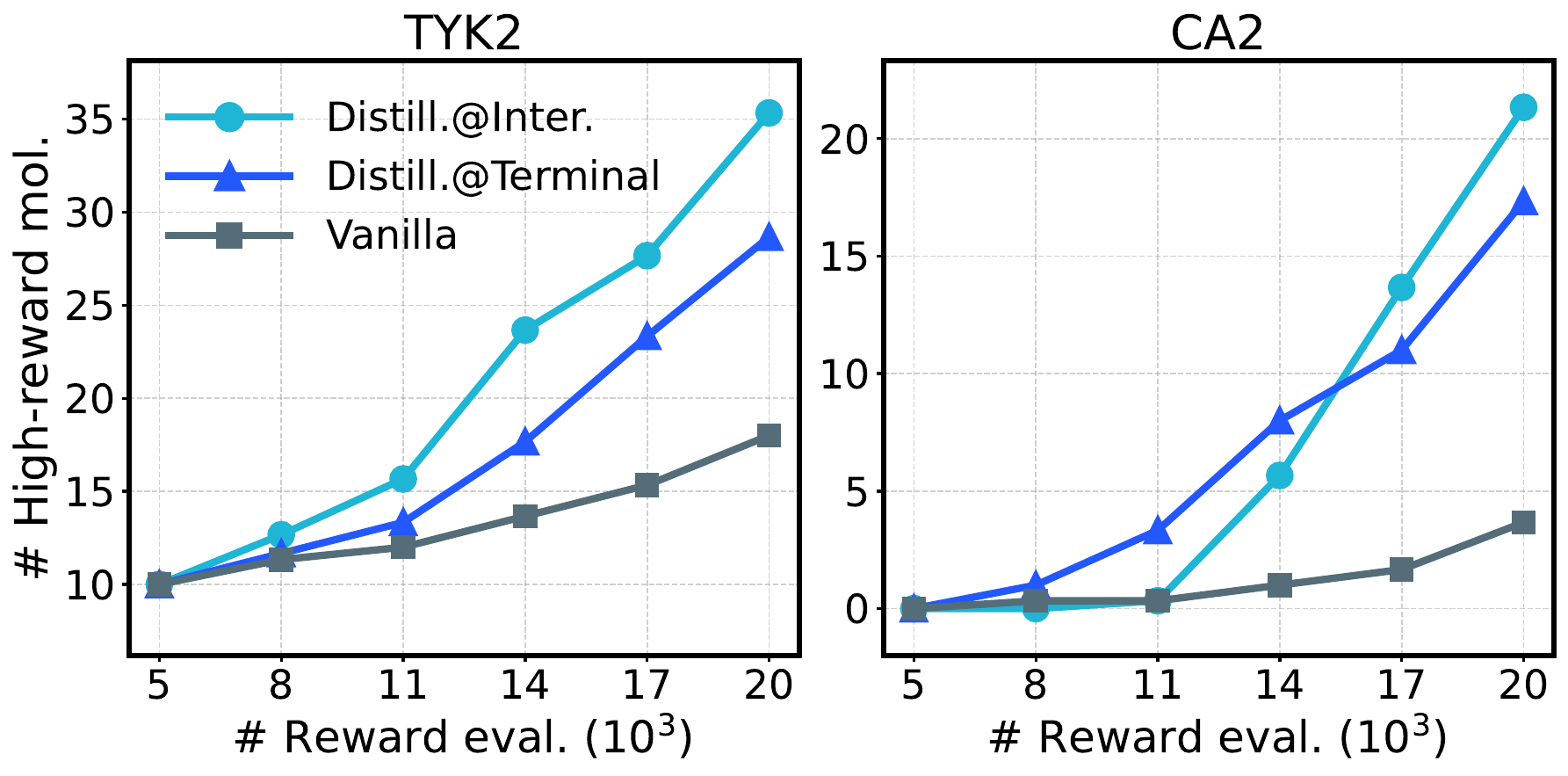}
\caption{\textbf{Representation alignment on intermediate molecules.}}\label{fig:distill_compare}
\end{figure}

We further conduct an ablation study that extends representation alignment-based distillation from generated molecules to intermediate molecules produced by the policy before generation. Specifically, we consider the following alignment loss:
\begin{equation*}
\mathcal{L}^{\text{inter.}}_{\text{align.}}(\tau=(s_0,\ldots,s_T),f(m))=
-\lambda\cdot\sum^{T}_{t=0}\cos\!\big(
h_{\theta}(s_t),
f(m)_{\mathcal{I}(s_t)}
\big),
\end{equation*}
where $\mathcal{I}(s_t)$ denotes the index set of atoms in the terminal molecule $s_T$ that correspond to the molecular substructure present at the intermediate state $s_t$, and $f(m)_{\mathcal{I}(s_t)}$ denotes the representations of this corresponding substructure. An intermediate molecule $s_t$ may lead to multiple terminal molecules through different trajectory actions, resulting in multiple possible alignment targets. To address this, we restrict alignment to representations of terminal molecules in the top $10\%$ by reward. We report the results of this \Cref{fig:distill_compare}

\newpage

\newpage

\section{Ablations on pooling}\label[appendix]{supp:pooling_ablation}

\begin{table}[h]
\caption{\textbf{Ablations on pooling strategy.} \textbf{Bold} indicates the best performance. Boltz2 consistently shows strong performance compared to the baselines without advanced pooling strategy.}
\label{tab:ablation_pooling}
\centering
\begin{tabular}{l c c c c c c c c}
\toprule
\multicolumn{1}{@{}c@{}}{} &
\multicolumn{4}{@{}c@{}}{Baselines} &
\multicolumn{4}{@{}c@{}}{(Ours) Pooling w/o} \\
\cmidrule(lr){2-5}
\cmidrule(lr){6-9}
\textbf{Tasks} &
\textbf{MolE} &
\textbf{KGPT} &
\textbf{Mini.} &
\textbf{QIP} &
\textbf{Bond} &
\textbf{Diag} &
\textbf{Std} &
\textbf{$\varnothing$} \\
\midrule
Solubility $\downarrow$ & 0.79 & 0.71 & 0.74 & 0.70 & 0.66 & 0.66 & \textbf{0.64} & 0.66 \\
BBB $\uparrow$ & 0.90 & 0.91 & 0.92 & 0.90 & 0.92 & 0.92 & 0.92 & \textbf{0.93}  \\
CYP2C9 V.\ $\uparrow$ & 0.80 & 0.80 & \textbf{0.82} & 0.78 & \textbf{0.82} & \textbf{0.82} & \textbf{0.82} & \textbf{0.82} \\
Half Life $\uparrow$ & 0.55 & 0.53 & 0.50 & 0.53 & 0.61 & 0.63 & \textbf{0.64} & 0.62 \\
LD50 $\downarrow$ & 0.82 & 0.55 & 0.59 & 0.56 & \textbf{0.40} & \textbf{0.40} & 0.41 & \textbf{0.40} \\
\bottomrule
\end{tabular}
\end{table}

We conduct an ablation study on pooling strategy in Supplementary Table~\ref{tab:ablation_pooling}. Our hybrid pooling captures diverse structural statistics of pair representations, it yields only marginal improvements over simpler pooling schemes. Importantly, even with reduced pooling, Boltz representations generally outperform existing molecular representation models.

\newpage



\subsection{Boltz2 modification} 
We modify the official Boltz2 code \citep{passaro2025boltz2}, which is released under the MIT license, to operate in a ligand-only setting. The modification is as follows:
\begin{itemize}
\item Boltz2 yields errors when the input protein sequence is empty. We therefore use a single \texttt{X} token as the protein sequence input and modify the Pairformer trunk to ignore it by slicing the corresponding indices.
\item Boltz2 uses 3D conformations computed with the Universal Force Field (UFF) implemented in RDKit. For molecules whose 3D conformations cannot be initialized with UFF, we initialize conformations by sampling from $[-1, 1]$. 
\end{itemize}
In the code, we include the modified Boltz2 implementation used in our work.


\end{document}